\documentclass[a4paper]{article}
\usepackage{RRA4}
\usepackage{amssymb}
\usepackage{amsmath}
\usepackage[caption=false,font=footnotesize]{subfig}
\newtheorem{ass}{Assumption}
\newtheorem{theorem}{Theorem}
\newenvironment{proof}{\noindent \textbf{Proof:
}}{\hspace{\stretch{1}}$\square$}
\newenvironment{soproof}{\noindent \textbf{Sketch of proof:}}{\hspace{\stretch{1}}$\square$}

\long\def\jump#1\finjump{}

\newcommand{\ddx}{\, \mathrm{d}x}
\newcommand{\atoi}[1]{\lfloor #1 \rfloor+1}

\newcommand{\dd}{\mathrm{d}}
\newcommand{\ed}{\mathrm{e}}
\newcommand{\prob}{\mathbb{P}}
\newcommand{\mS}{\mathcal{S}}
\newcommand{\mD}{\mathcal{D}}
\newcommand{\mG}{\mathcal{G}}
\newcommand{\ms}{\mathcal{S}}
\newcommand{\md}{\mathcal{D}}

\RRdate{September 2008}
\RRauthor{
Fabien Mathieu\thanks{Orange Labs}
\and
Gheorghe Postelnicu\thanks{Google}
\and
Julien Reynier\thanks{Reech AiM Research}%
}
\authorhead{Mathieu \& al.}
\RRtitle{Configuration stable des systèmes à préférences acycliques}
\RRetitle{The stable configuration in\\ acyclic preference-based systems}
\titlehead{Stable configuration of acyclic systems}
\RRnote{Supported by CRC MARDI II}
\RRnote{Supported by ANR Project Shaman}
\RRresume{
Les systèmes à préférences acycliques sont récemment apparus comme une méthode élégante de modélisation de certains systèmes ditribués de type pair-à-pair. Une instance acyclique admet une unique configuration stable, auto-stabilisante, qui donne une bonne indication du comportement du système. Dans ce rapport, nous donnons la distribution statistique de la configuration stable pour trois types de préférences acycliques : les préférences globales (basées sur un ordre total des n\oe uds), les préférences de distance (le plus proche est préféré), et les préférences acycliques aléatoires. Sous l'hypothèse d'un graphe de compatibilité Erdös-Rényi, nous montrons à l'aide de techniques de limites fluides et de champ moyen l'existence d'une distribution limite continue. La pertinence des résultats est vérifiée à l'aide de simulations.
}
\RRabstract{
Acyclic preferences recently appeared as an elegant way to model many distributed systems. An acyclic instance admits a unique stable configuration, which can reveal the performance of the system. In this paper, we give the statistical properties of the stable configuration for three classes of acyclic preferences: node-based preferences, distance-based preferences, and random acyclic systems. 
Using random overlay graphs, we prove using mean-field and fluid-limit techniques that these systems have an asymptotically continuous independent rank distribution for a proper scaling, and the analytical solution is compared to simulations.
These results provide a theoretical ground for validating the performance of bandwidth-based or proximity-based unstructured systems.
}
\RRmotcle{Systèmes acycliques, distribution, limite fluid, champ moyen, petit-mondes, EDP}
\RRkeyword{Acyclicity, rank distribution, fluid limit, mean-field, small-worlds, PDE}
\RRprojets{GANG}
\RRtheme{\THNum} 
\URRocq 
\usepackage{url}
\usepackage{hyperref}

\begin{document}
\RRNo{6628}
\makeRR   

\tableofcontents
\newpage

\section{Introduction}
\label{sec:introduction}

Matching problems with preferences have applications in a variety of real-world situations, including dating agencies, college admissions, roommate attributions, assignment of graduating medical students to their first hospital appointment, or kidney exchanges programs~\cite{gale62college,irving00hospitals,irving02stable,roth84evolution,roth05pairwise}. 

Recently, matching problems also appeared as an elegant way to model many distributed systems, including ad-hoc and peer-to-peer networks~\cite{lebedev07using,DBLP:conf/europar/GaiLMMRV07,mathieu08self,DBLP:conf/sss/Mathieu07,DBLP:conf/icdcs/GaiMMR07}. In distributed systems, the preferences generally come from direct measurements. Those measurements can be node-related (CPU, upload/download bandwidths, storage, battery, uptime), or edge-related (Round-Trip Time, physical/virtual distances, link capacity, co-uptime). In most cases, the resulting preferences are acyclic: there cannot exist a cycle of more than two nodes such that each node prefers its successor to its predecessor. As a consequence, there always exists a unique stable configuration, which is self-stabilizing~\cite{DBLP:conf/europar/GaiLMMRV07,DBLP:conf/wine/AbrahamLMO07}. This makes things much easier than in other matching problems, where finding, counting and comparing the stable configurations are some of the main issues~\cite{gale62college,tan91necessary,mertens05random,roth05pairwise}.

Modeling distributed systems with acyclic preferences allows us to predict the effective collaborations that will occur, which, in turn, allows us to infer the performance of a given system. For convenience, the study of an acyclic distributed system is often split into two main problems:
\begin{itemize}
	\item \emph{How fast is the stabilization process?} Because distributed systems are often highly dynamic, with constant churn and preference alteration, the speed of convergence can be used to determine how far the effective configurations are from the time-evolving stable configuration.
	\item \emph{What are the properties of the stable configuration?} If the stabilization process is fast enough, the effective and stable configurations will be close. Analyzing the latter can then give valuable information on the former.
\end{itemize}

In a previous work, Mathieu investigated the first question~\cite{mathieu08self,DBLP:conf/sss/Mathieu07}. He proved that even if the convergence can be prohibitive under an adversary scheduler, it is fast for realistic scenarios. The second question has been answered for specific acyclic preferences: for real-world latency-based preferences, the stable configuration shows, for $b$-matching (several mates per nodes allowed), small-world properties (low diameter and high clustering coefficient)~\cite{DBLP:conf/europar/GaiLMMRV07};
for node-related preferences, the stable configuration tends to pair nodes with similar values~\cite{DBLP:conf/icdcs/GaiMMR07}: this is the stratification effect, which allows, for instance, to understand upload/download correlations in incentive networks like BitTorrent~\cite{cohen03incentives}.

\subsection{Contribution}
The studies proposed in~\cite{DBLP:conf/icdcs/GaiMMR07} and~\cite{DBLP:conf/europar/GaiLMMRV07} gave only partial, mostly empirical, answers about the link distribution in the stable configuration, and proposed some conjectures. The goal of this paper is to complete and give theoretical proofs on the shape of the stable configuration.

We extend the seminal results that were given in~\cite{DBLP:conf/icdcs/GaiMMR07} for node-based preferences: for $b=1$ (simple matching case), we prove the existence of a limit continuous distribution and solve the corresponding Partial Differential Equation (PDE). Then we apply a similar method for distance-based and random-acyclic preferences, and also give the explicit solution of the corresponding PDE.

Lastly, we extend the results for $b>1$ (multiple matchings). In that case, there is no simple expression that gives the exact solution of the PDEs system, but discrete equations are used to observe asymptotical behavior of the distribution. For node-based preferences, the exponential behavior validates the stratification effect (the probability to be matched with a distant peer decreases exponentially with the distance), while the power law obtained for the two other cases indicates that the small world effect observed in~\cite{DBLP:conf/europar/GaiLMMRV07} for latency is in fact common to all distance-based preferences\footnote{Latencies cannot be considered as real distances, mainly because the triangular inequality is not always verified. However, they form an inframetric, which is no too far from a real metric~\cite{flv08}.}.

\subsection{Roadmap}
In Section~\ref{sec:model} we define the model and notation for preference-based systems. Section~\ref{sec:mean_field} gives the generic mean field method used in this paper to solve the simple matching case. The case of node-based preferences is solved in Section~\ref{sec:nb}, then the results are adapted to the distance-based and random-acyclic preferences in Section~\ref{section:ra_db}. Section~\ref{sec:bb_mean} extend the formulas to multiple matchings, and asymptotical properties of the distributions are described. Lastly, Section~\ref{sec:conclusion} concludes.

\section{Model and notation}
\label{sec:model}

A preference-based system is a set $V$ of $N$ nodes, whose possible interactions are described by an acceptance graph $G$, a mark matrix $m$ and a quota vector $b$.

The quota vector $b$ limits the collaborations: a peer $i$ cannot have more $b(i)$ simultaneous mates.

The acceptance graph $G=(V,E)$ is an undirected, non-reflexive graph. It describes allowed matchings: a node $i$ and a node $j$ can be mated (we say that $i$ is acceptable for $j$, and \emph{vice versa}) if, and only if (iff) $\{i,j\}\in E$. For instance, in peer-to-peer networks, a node cannot be directly connected to all other peers of the system, because of scalability, and peers that are not directly connected cannot be mated. In this paper, we consider Erdös-Rényi graphs $\mG(N,p)$ (each possible edge exists with probability $p$ independently of the others; hence the expected degree is $d=p(N-1)$).

The mark matrix $m$ is used to construct the peers' preferences: given two nodes $j$ and $k$ acceptable for $i$, $i$ ranks $j$ better than $k$ iff $m_{i,j}<m_{i,k}$ (the sign is arbitrary). The following marks are considered in this paper:
	\paragraph{Node-based} $m(.,i)$ is constant (nodes have intrinsic values). These preferences are suited to modeling peer-related performance, like access bandwidth, storage, CPU, uptime\ldots
	\paragraph{Geometric} the nodes are associated to $N$ points picked uniformly at random on a $n$-dimensional torus ($n\geq 1$). The marks are the distances between those points. These preferences allow a theoretical analysis of proximity-based performance.
	\paragraph{Meridian latencies} we considered random subsets of $N$ nodes taken from the $2500$ nodes dataset of the Meridian Project~\cite{meridianproject}. The marks are the (symmetric) latencies between those nodes. We do not perform analysis for those marks, but use them in \S~\ref{subsec:diaclust} for validating the geometric approach.
	\paragraph{Random acyclic} each edge receives a random uniformly distributed value. The name is justified because all acyclic preferences can be described by marks on the edges (which is equivalent to assume that $m$ is symmetric). Hence uniformly distributed (symmetric) random marks are a convenient way to perform a uniform sampling of the acyclic preferences~\cite{DBLP:conf/europar/GaiLMMRV07,DBLP:conf/wine/AbrahamLMO07}.

All the considered marks are acyclic, and therefore a $(G,m,b)$ system admits a unique stable configuration $C\in E$, which is self-stabilizing~\cite{lebedev07using,mathieu08self}. The neighbors of $i$ in $C$ are the stable mates of $i$, and the notation $i \leftrightarrow j$ is used to express that $i$ and $j$ are stable mates.

 We assume for simplicity that $m$ is complete and not limited to the edges of $\mG$. For all considered preferences but random acyclic, the completion is straightforward. For random acyclic preferences, we assume that dummy random values are assigned to non-acceptable edges.

The preferences are denoted like follows: if $j$ is acceptable for $i$, $r_i(j)$ denotes the rank of $j$ in $i$'s list ($1$ being the best). $r_i$ is called the \emph{acceptable} ranking of $i$. If  $i$ has more than $k$ acceptable neighbors, $r_i^{-1}(k)$ is the $k^{th}$ node in $i$'s acceptable ranking. Similarly, for $j\neq i$, $R_i(j)$ denotes the rank of $j$ in the complete graph (the acceptability condition is omitted). $R_i$ is called the \emph{complete} ranking of $i$ For $K<N$, $R_i^{-1}(K)$ is the $K^{th}$ node in $i$'s complete ranking.

All stable mating probabilities that are discussed in this article are designed by $D$. Subscripts and arguments are used to precise the meaning of $D$ whenever needed. For instance:
\begin{itemize}
	\item $D_{R_i}(K)$ is the probability that $i$ has a stable mate with complete rank $K$.
	\item $D_{N,d}(i,j)$ is the probability that $i\leftrightarrow j$, knowing there is $N$ nodes and that the expected degree of the acceptance graph is $d$.
	\item for $c\leq b(i,)$, $D_{r_i,c}(k)$ is the probability that the $c^{th}$ stable mate of $i$ has relative rank $k$.
	\item \ldots
\end{itemize}
The complementary cumulative distribution function (CCDF) of $D$ is denoted $S$, and the scaled version of $D$ and $S$ are denoted $\mD$ and $\mS$.

\section{Acyclic formulas}
\label{sec:mean_field}

We first consider the case $b=1$ (simple matching) (the results will be extended to multiple matchings in Section~\ref{sec:bb_mean}). We give a generic formula that describes the complete rank of the mate $C(i)$ of a peer $i$.

\subsection{Generic formula}
\label{subsec:generic}

Let $D_{R_i}(K)$ be the probability that $R_i(C(i))=K$ (the probability that the mate of $i$, if any, has rank $K$). The CCDF of $D$ is  $S_{R_i}(K):=1-\sum_{L=1}^{K-1}$, which is the probability that $i$'s mate has a rank greater than $K$ ($R_i(C(i))\geq K$) or has no mate (short notation: $R_i(C(i))\nless K$). Following the approach proposed in~\cite{DBLP:conf/icdcs/GaiMMR07}, we first give a generic exact formula that describes $D_{R_i}$, then we propose a simplified mean-field approximation.

In order to solve $D_{R_i}(K)$, one can observe that $i$ is mated with its $K^{th}$ peer $j=R^{-1}_i(K)$ iff:
\begin{itemize}
	\item $\{i,j\}$ is an edge of the acceptance graph; this happens with probability $p$ as $G$ is supposed to be a $\mG(N,p)$ graph.
	\item $i$ is not mated with a node better than $j$ ($R_i(C(i))\nless K$);
	\item $j$ is not mated with a node better than $i$ ($R_j(C(j))\nless R_j(i)$).
\end{itemize}

This leads to the following exact formula:
\begin{equation}
\begin{array}{rl}
D_{R_i}(K) & =  p\prob(R_i(C(i))\nless K)\times \\
&  \qquad \times \prob(R_j(C(j))\nless R_j(i)|R_i(C(i))\nless K)\\
& =  pS_{R_i}(K)\prob(R_j(C(j))\nless R_j(i)|R_i(C(i))\nless K)
\end{array}
	\label{eq:d_exact}
\end{equation}

\subsection{Mean-field approximation}
\label{subsec:meanfield}

Solving~\eqref{eq:d_exact} is difficult to handle, mainly because of possible correlations between $R_j(C(j))\nless R_j(i)$ and $R_i(C(i))\nless K$. The solution is to adopt a mean field assumption:

\begin{ass} The events \emph{node $i$ is not with a node better than $j$} and \emph{node $j$ is not with a node better than $i$} are independent.
\label{ass:meanfield1}
\end{ass}

This assumption has been proposed in~\cite{DBLP:conf/icdcs/GaiMMR07} to solve~\eqref{eq:d_exact} in the case of node-based preferences. It is reasonable when $N$ is large and $p$ is small. Then \eqref{eq:d_exact} can be approximated by
\begin{equation}
D_{R_i}(K) = pS_{R_i}(K) S_{R_j}(R_j(i))\text{.}
		\label{eq:d_approx0}
\end{equation}

Now, in the next two sections, we propose to solve Equation~\ref{eq:d_approx0} for specific preferences.

\section{Node-based preferences}
\label{sec:nb}

We assume here that the preferences comes from marks on nodes.
This is equivalent to assume a total order among the nodes. Therefore we do not need to explicit the mark matrix $m$, and we can use an ordered node labeling instead. We arbitrary choose $1,\ldots,N$ as labels, $1$ been the best (if $1$ is ranked first for all nodes that accept $1$, and so on\ldots).

Because the nodes' label express their complete ranks, we can directly consider $D(i,j)$, the probability that node $i$ is mated with node $j$. Node $j$ has rank $j$ for $i$ if $j<i$, and $j-1$ if $j>i$, because a rank does not rank itself.  This gives the relation between $D$ and $D_{R}$:
\begin{equation}
D(i,j) = \left\{
\begin{array}{l}
	D_{R_i}(j)\text{ if $j<i$,}\\
	0 \text{ if $j=i$ (mating is not reflexive),}\\
	D_{R_i}(j-1)\text{ if $j>i$.}
\end{array}
\right.
\label{eq:dij_dri}
\end{equation}

Using the CCDF $S(i,j):=1-\sum_{k=1}^{j-1}D(i,k)$, we get the node-based version of Equation~\ref{eq:d_approx0}:
\begin{equation}
D(i,j)=\left\{
\begin{array}{l}
0	\text{ if $i=j$,}\\
pS(i,j)S(j,i)	\text{ otherwise.}
\end{array}
\right.
	\label{eq:dij_sij}
\end{equation}
This equation, which was originally proposed in~\cite{DBLP:conf/icdcs/GaiMMR07}, which also show that it gives a very good approximation of empirical distribution. It can be numerically solved by using a double iteration.

\subsection{Fluid limit}

Our main contribution for node-based preferences is to prove that, under a constant degree scaling, $D$ admits a fluid limit. This limit gives a complete description of $D$ that can be applied to all values of $N$ and $p$, while Equation~\eqref{eq:dij_sij} needs to be solved for each set of parameters.

\subsection{Constant degree scaling}

In order to compare the distributions for arbitrary values of $N$, we need a scaled version of $D$, where a peer $i$ is represented by a scaled ranking $0\leq \alpha <1$. In details, we associate to each $i$ the number $\alpha(i)=\frac{i-1}{N}$, and to each real number $alpha$ the node $i(\alpha)=\atoi{N\alpha}$. The scaled version of $D$, denoted $\mD$, is then defined by $$\mD_N(\alpha,\beta)=ND(\atoi{N\alpha},\atoi{N\beta})\text{.}$$

$\mD_N$ is a piecewise constant function. Its set of function values is the set of the $(ND(i,j))$ values. The factor $N$ in its definition allows to express $D(i,j)$ as an integral of $\mD$:
$$D(i,j)=\int_{\frac{j-1}{N}}^\frac{j}{N}\mD_N(\frac{i-1}{N},x)\, \mathrm{d}x=\int_{\frac{i-1}{N}}^\frac{i}{N}\mD_N(x,\frac{j-1}{N})\, \mathrm{d}x$$

The scaling of the CCDF is defined by 
\begin{equation}
	\mS_N(\alpha,\beta)=1-\int_{0}^{\beta}\mD_N(\alpha,x)\, \mathrm{d}x\text{,}
	\label{eq:sab_def}
\end{equation}
and the relation between $S$ and $\mS$ is
\begin{equation}
S(i,j)=\mS(\frac{i-1}{N},\frac{j-1}{N})\text{.}	
	\label{eq:sij_sab}
\end{equation}

\subsection{Convergence theorem}

We now want to show the existence of a continuous limit for $\mD$. The problem is the existence of a discontinuity for $\alpha\approx\beta$, because $D(i,i)=0$. However, this discontinuity is just a reminder of the fact that a node cannot mate with itself, so we propose to make $\mD$ more ``continuous'' by introducing
$$\tilde{\mD}(\alpha,\beta)=\left\{
\begin{array}{l}
	\mD(\alpha,\beta)\text{ if $\lfloor N\alpha \rfloor\neq\lfloor N\beta\rfloor$,}\\
	Np(S(\atoi{N\alpha},\atoi{N\alpha}))^2 \text{ otherwise.}
\end{array}
\right.$$

The fluid limit of $\tilde{\mD}$ is then given by the following theorem:
\begin{theorem}
Let $d>0$ be a constant. If $N\rightarrow \infty$ with $p=\frac{d}{N}$, the function $\tilde{\mD}_{N,d}$ uniformly converge towards 
\begin{equation}
\mD_\infty(\alpha,\beta)=\frac{de^{d(|\beta-\alpha|)}}{(1-e^{-d\min(\alpha,\beta)}+e^{d|\beta-\alpha|})^2}\text{.}
	\label{eq:f_nb_d}
\end{equation}
\label{th:f_nb_d}
\end{theorem}

This result indicates that asymptotically, the average degree in the acceptance graph completely defines the mating distribution. The consequence is that we can explicitly describe the so-called \emph{stratification} effect~\cite{DBLP:conf/icdcs/GaiMMR07}:
the mating distribution is exponentially decreasing with $|\beta-\alpha|$, with intensity $d$. In other words, a peer with scaled rank $\alpha$ tends to mate with a mate of same scaled rank, with a standard deviation of the same order than $\frac{1}{d}$.

The proof of Theorem~\ref{th:f_nb_d} is given in Appendix~\ref{app:equ_node}. Note, that the existence of a fluid limit was proposed as a conjecture in~\cite{DBLP:conf/icdcs/GaiMMR07}, and proved for $\alpha=0$ (but the expression of the fluid limit in the general case was not provided).

Theorem~\ref{th:f_nb_d} gives two corollaries:
\begin{itemize}
	\item using the CCDF of $\mD_\infty$, the probability that a node of scaled rank $\alpha$ has no mate is $\frac{1}{1+e^{-d\alpha}(e^{-d}-1)}$;
	\item  for $i\neq j$ (discrete case), a good approximation for $D(i,j)$ is
	\begin{equation}	D(i,j)\approx\frac{pe^{p(|j-i|)}}{(1-e^{-p\min(i,j)}+e^{p|j-i|})^2}\text{.}
	\label{eq:dij_explicit}
\end{equation}
\end{itemize}

\subsection{Validation}

We compared our fluid limit approximation, given by \eqref{eq:dij_explicit}, to the mean-field  values given by~\eqref{eq:dij_sij}, which are known to be accurate (\cite{DBLP:conf/icdcs/GaiMMR07}).

\begin{figure}[ht]
\begin{center}
\subfloat[$N=50$,$d=5$,$\alpha=\frac{1}{10}$]{\includegraphics[width=.4\textwidth]{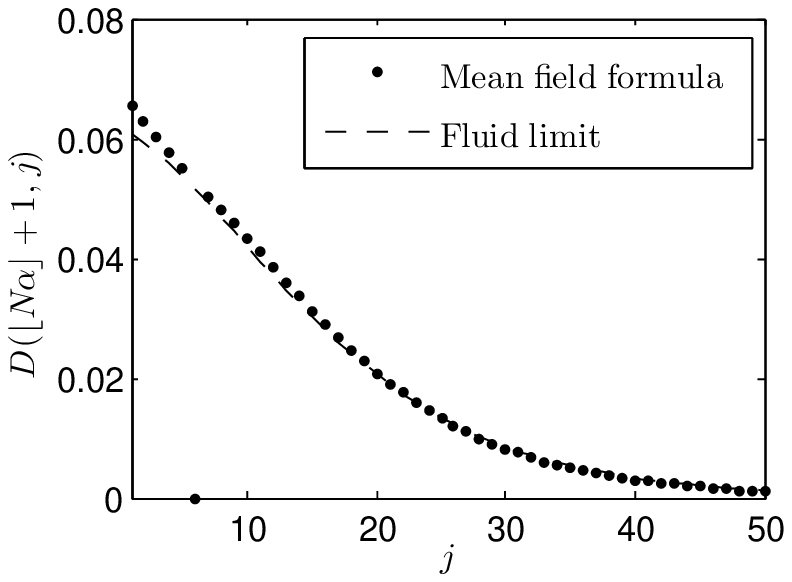} \label{fig:nodebased_b_1_N_50_d_5_i_6}}
\subfloat[$N=50$,$d=30$,$\alpha=\frac{9}{10}$]{\includegraphics[width=.4\textwidth]{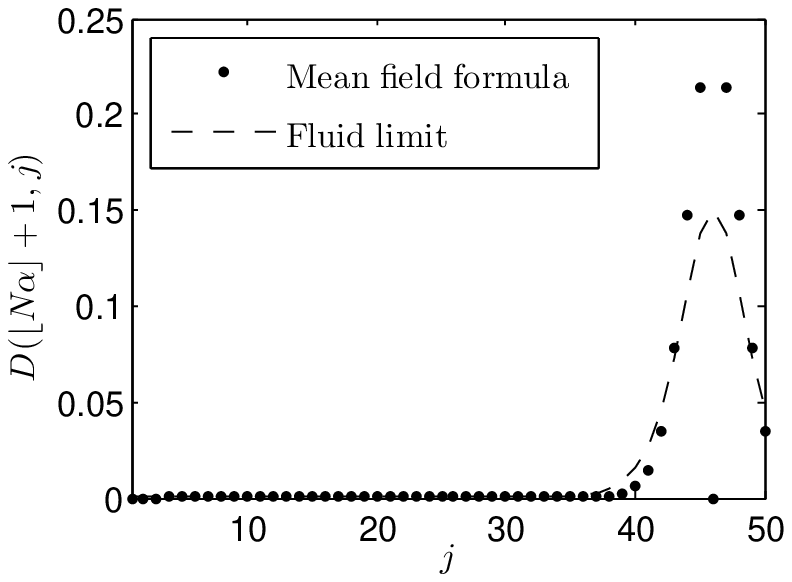} \label{fig:nodebased_b_1_N_50_d_30_i_46}}\\


\subfloat[$N=2000$,$d=5$,$\alpha=\frac{1}{10}$ ]{\includegraphics[width=.4\textwidth]{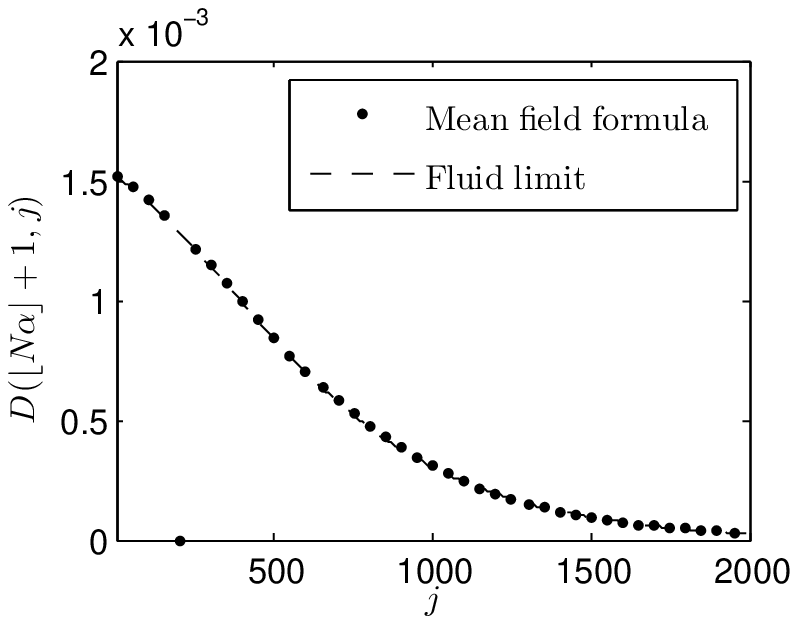} \label{fig:nodebased_b_1_N_2000_d_5_i_201}}
\subfloat[$N=2000$,$d=30$,$\alpha=\frac{9}{10}$ ]{\includegraphics[width=.4\textwidth]{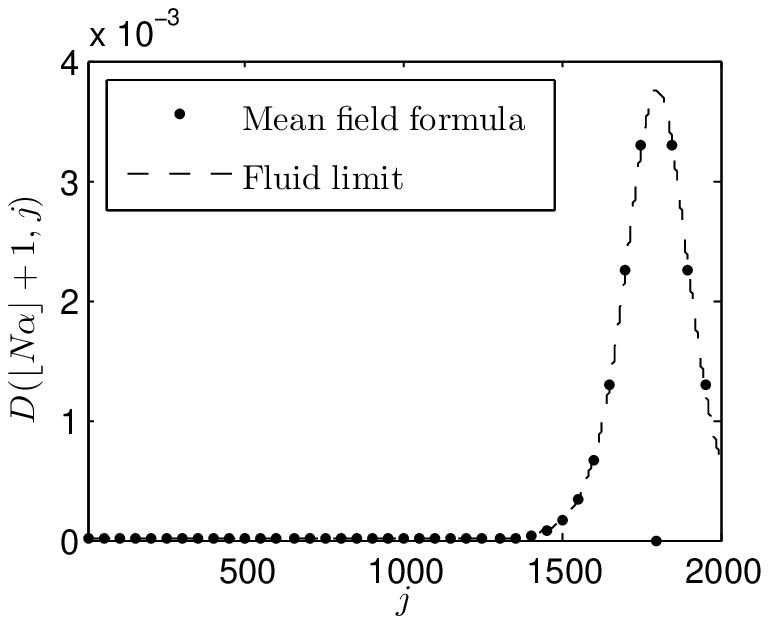} \label{fig:nodebased_b_1_N_2000_d_30_i_1801}}
	\caption{Validation of the fluid limit for node-based preferences.}
	\label{fig:nodebased_valid}
\end{center}
\end{figure}

$N$ was set to $50$ or $2000$, and $d$ to $5$ or $30$. Because $D$ is $2$-dimensional, we arbitrary set the scaled rank $\alpha$ to $0.1$ or $0.9$ (but the convergence validation holds for any $\alpha$).
 The results are shown in Figure~\ref{fig:nodebased_valid}.

We observe a gap for $j=\lfloor N\alpha\rfloor+1$, because the mean field formula sets $D$ to $0$ whereas the fluid limit uses a continuous extension.

Besides this gap, $N=50$ (Figures \ref{fig:nodebased_b_1_N_50_d_5_i_6} and \ref{fig:nodebased_b_1_N_50_d_30_i_46}) shows some difference between the mean field and the fluid limit. The error is especially noticeable for $d=30$ (\ref{fig:nodebased_b_1_N_50_d_30_i_46}). However
For $N=2000$ (Figures \ref{fig:nodebased_b_1_N_2000_d_5_i_201} and \ref{fig:nodebased_b_1_N_2000_d_30_i_1801}), there is practically no error.

These results are consistent with~\eqref{eq:bound_for_E} (in the Appendix~\ref{app:equ_node}), which shows that the convergence is $O(\frac{d^2}{N}e^{8d})$.

\subsection{Exact resolution}

For the record, if $b=1$, there exists an exact recursive formula for the node-based stable configuration. This formula is 
\begin{equation}
\begin{array}{rl}
D(i,j)= & \left(1-S(1,i)\right)D(i-2,j-2)\\
&+\left(S(1,i+1)-S(1,j)\right)D(i-1,j-2)\\
&+S(1,j+1)D(i-1,j-1)
\end{array}	\text{ for $i<j$,}
	\label{eq:dij_exact_in}
\end{equation}
with the border conditions $D(1,k)=p(1-p)^{k-2}$ ($k\geq 2$), $D(i,i)=0$.

This equation also admits a fluid limit, which happens to be the same than the fluid limit of the mean field formula. This result appears as a strong validation of the mean field approach: although the mean field formula is not exact (its results differ from the exact formula), its fluid limit is exact.

One could wonder why using a mean field formula if a usable exact formula exists. The issue with the exact formula is that it relies on a ``trick'': if you remove node $1$ and its mate from the system, the remaining nodes still form a preference-based system with same parameters except there is two less nodes. However, this trick cannot be generalized for other preferences or for $b>1$. This is why we focus on the mean field formulas.

A complete proof of the exact recursive formula, including its PDE counterpart and resolution, can be found in Appendix~\ref{sec:b1_exact}.


\section{Acyclic and distance-based preferences}
\label{section:ra_db}

We now consider geometric and random acyclic preferences. Following the approach used for node-based preferences, we first focus on the complete rank distribution. Use mean field assumptions, we propose a recursive formula for $D$, then we solve the fluid limit. The results are then extended to the distance and acceptable rankings distributions.

\subsection{Complete rank distribution}
\label{section:rank}

Assumption~\ref{ass:meanfield1} is not enough for solving~\eqref{eq:d_approx0} in the case of geometric or random acyclic preferences. Therefore, we propose this additional assumption:
\begin{ass} For geometric and random acyclic preferences, the following approximations hold:
\begin{itemize}
	\item $D_{R_i}(K)$ is independent of $i$ (and therefore denoted $D_R(K)$);
	\item the complete ranking is symmetric: $R_i(j)=R_j(i)$.
\end{itemize}
\label{ass:meanfield2}
\end{ass}

The first approximation just states that in average, all nodes have the same mate distribution, while the second one tells that $R_i(j)$ is a good approximation of $R_j(i)$. These approximations were motivated by the uniform distributions used for shaping the preferences. In particular, they do not apply for node-based preferences, where the mate distribution is strongly affected by a node's mark. Under these assumptions, we get
\begin{equation}
	D_R(K)=pS^2_R(K)\text{, with $S_R(K)=1-\sum_{L=1}^{K-1}D_R(L)$.}
	\label{eq:d_approx}
\end{equation}

This equation gives an immediate recursion for $S_R$:
\begin{equation}
S_R(K)=	\left\{\begin{array}{l}
	1\text{ if $K=1$,}\\
	S_R(K-1)-pS_R^2(K-1)\text{ otherwise.}
\end{array} \right.
	\label{eq:s_approx}
\end{equation}

In return, $D_R$ is directly given by $D_R(K)=S_R(K)-S_R(K+1)$.

\subsubsection{Fluid limit}
\label{subsec:fluid}

We now give the fluid limit of $D_R$. The scaled version of $D_R$ is defined like for node-based preferences, except that there now only one parameter. For $0\leq\alpha<1$, we define $\md_R(\alpha):=(N-1)D_R(\atoi{(N-1)\alpha\rfloor})$. The scaling factor is now $N-1$ because it is the upper bound for $K$ (while $N$ was the upper bound for $i,j$ in~\S\ref{sec:nb}). $D_R$ can be expressed as an integral of $\md_R$: $D_R(K)=\int_{\frac{K-1}{N-1}}^{\frac{K}{N-1}}\md_R(x)\ddx$. The scaled CCDF, $\ms_R$, is then naturally defined as:
$$\ms_R(\alpha)=1-\int_{0}^{\alpha}\md_R(x)\ddx\text{.}$$

\begin{theorem}
We assume that $d=p(N-1)$ is a positive constant. As $N\rightarrow \infty$, $\mS_R$ uniformly converges towards 
\begin{equation}
	\ms_\infty(\alpha)=\frac{1}{d\alpha+1}\text{.}
	\label{eq:f_sr_alpha}
\end{equation}

In particular, the probability that a node has no mate in the stable configuration is $\ms_R(1)=\frac{1}{d+1}$, and a good approximation for $S_R(K)$ is
\begin{equation}
	S_R(K)=\frac{1}{p(K-1)+1}\text{.}
	\label{eq:f_sr_k}
\end{equation}
\label{th:complete}
\end{theorem}

\begin{soproof}
The proof is a simpler version of the proof of Theorem~\ref{th:f_nb_d} (cf Appendix~\ref{app:equ_node}). First we prove that the $\mD_R$ functions are uniformly Cauchy (but in this case there is only one variable and there is no need for a continuous extension). This proves the uniform convergence towards $\mS_\infty$. Then we deduce from~\eqref{eq:s_approx} a differential equation verified by $\mS_\infty$:
\begin{equation}
	-\dot{\ms}_\infty(\alpha)=d\ms_R^2(\alpha)\text{,}
	\label{eq:fprime_c_alpha}
\end{equation}
\noindent with the boundary condition $\ms_R(0)=1$. The resolution of \eqref{eq:fprime_c_alpha} gives \eqref{eq:f_sr_alpha}, which completes the proof.
\end{soproof}

\subsubsection{Validation}

\begin{figure}[ht]
\begin{center}
\subfloat[$N=50$,$p=1$]{\includegraphics[width=.4\textwidth]{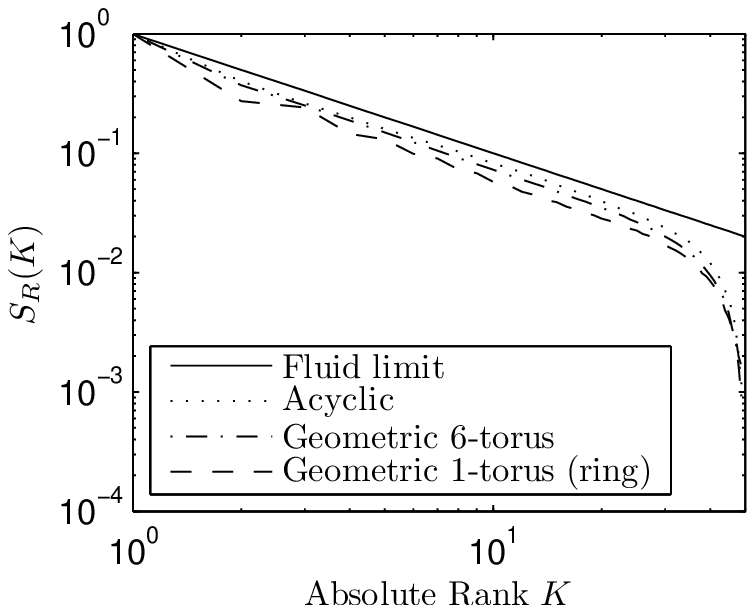} \label{eq:pvalid100_50}}
\subfloat[$N=2000$,$p=1$]{\includegraphics[width=.4\textwidth]{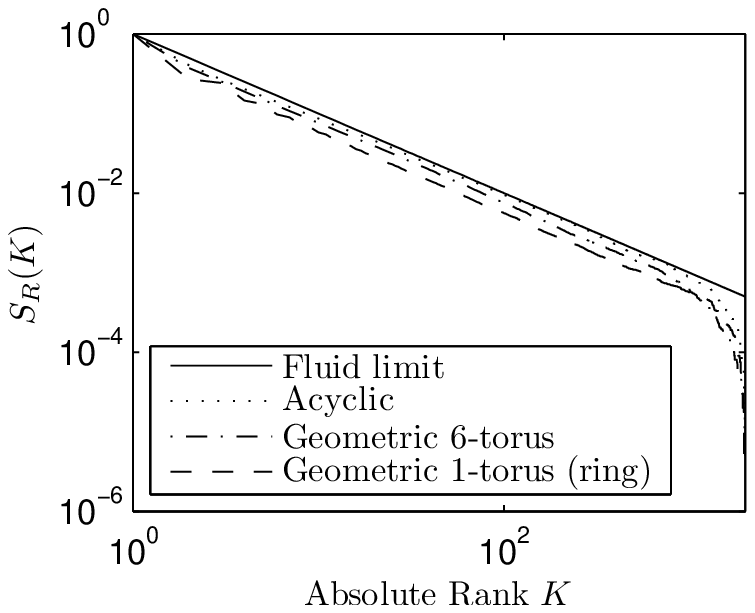} \label{eq:pvalid100}}\\
\subfloat[$N=50$,$p=\frac{1}{10}$]{\includegraphics[width=.4\textwidth]{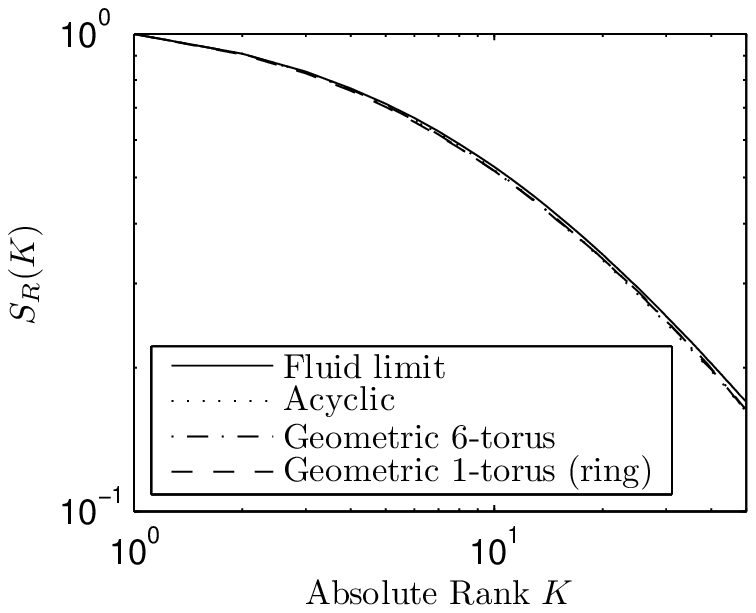} \label{eq:pvalid10_50}}
\subfloat[$N=2000$,$p=\frac{1}{10}$]{\includegraphics[width=.4\textwidth]{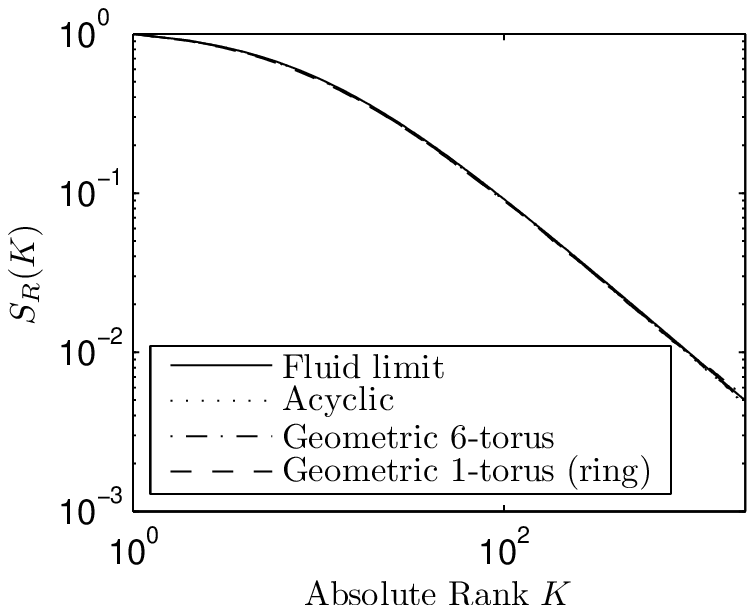} \label{eq:pvalid10}}\\
\subfloat[$N=50$,$p=\frac{1}{100}$]{\includegraphics[width=.4\textwidth]{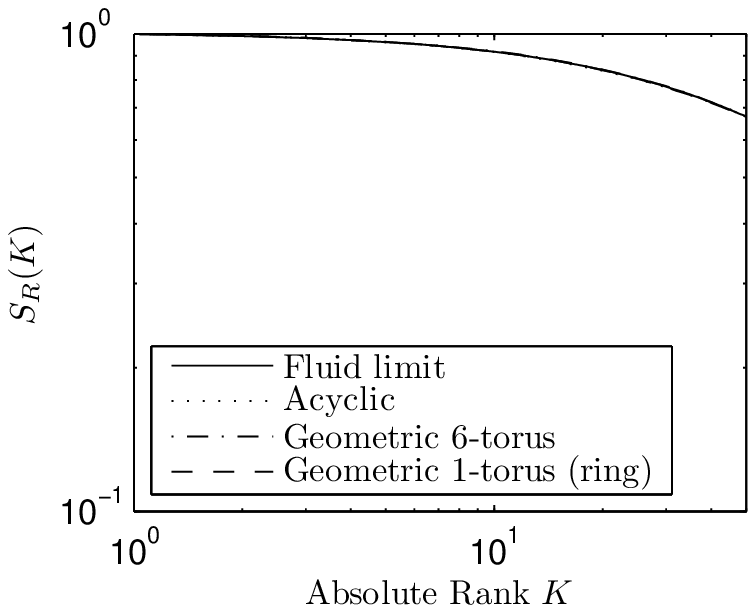} \label{eq:pvalid1_50}}
\subfloat[$N=2000$,$p=\frac{1}{100}$]{\includegraphics[width=.4\textwidth]{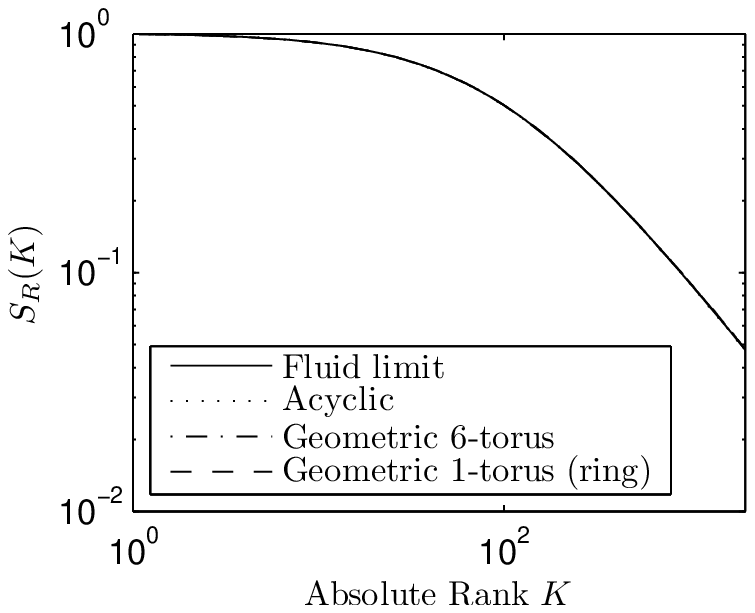} \label{eq:pvalid1}}

	\caption{Empirical validation of the fluid limit.}
	\label{fig:pvalid}
\end{center}
\end{figure}

Contrary to the case of node-based preferences, the mean field formula~\eqref{eq:d_approx} has not been validated in a previous work, so we could not compare the fluid limit with it, and used simulations\footnote{Actually, we did validate the mean field formula, but our results are to be published.}. We considered random acyclic instances, and geometric preferences in a $1$-dimensional torus and in a $6$-dimensional torus. $N$ was set to $50$ or $2000$. We used $3$ values of $p$: $1$, $\frac{1}{10}$ and $\frac{1}{100}$.  For each set of parameters, the empirical distribution was calculated over $100$ instances. The results are shown in Figure~\ref{fig:pvalid}.

For $p=1$ (Figures~\ref{eq:pvalid100_50} and~\ref{eq:pvalid100}), the mean-field assumptions hardly hold. As a consequence, the curves depend of the type of preferences, and the fluid limit is not accurate. This is especially visible if $K$ is close to the boundaries (that is $1$ or $N$). In particular, the non-mate probability is clearly over-estimated. However, the fluid limit manages to give the $O(\frac{1}{K})$ behavior that is common to all considered preferences. From that point of view, the fluid limit performs better than the recursive equation~\eqref{eq:s_approx}, which gives $S_R(K)=\delta_K^1$ for $p=1$.

For $p=\frac{1}{10}$ (Figures~\ref{eq:pvalid10_50} and~\ref{eq:pvalid10}), the curves are nearly indistinguishable. We verify that all types of preferences (acyclic or geometric) tend to have the same behavior and that Theorem~\ref{th:complete} gives precise approximations.

For $p=\frac{1}{100}$ (Figures~\ref{eq:pvalid1_50} and~\ref{eq:pvalid1}), the curves are indistinguishable.

We conclude that fluid-limit based on the mean-field formula is very effective for computing the complete ranking distribution, even if $N$ is not very large and $p$ is not very small.

\subsection{Distance distribution}
\label{subsec:geometric}

For geometric preferences, the actual distance between a node and its mate may be a more valuable performance indicator than the ranking. We call $S_X(x)$ the probability that the distance between a node $i$ and its mate $C(i)$ is not less than $x$ (in other words, the distance is greater than $x$ or $i$ is unmated). Under the fluid limit, we get a good estimate of $S_X$:
\begin{equation}
	S_X(x)=\frac{1}{dB_n(x)+1}\text{,}
	\label{eq:f_c_x}
\end{equation}
where $B_n$ is the size of a ball of radius $x$ in the $n$-torus.

\begin{proof}
In the fluid limit, a ball of radius $x$ contains $NB_n(x)$ nodes, because it occupies a ratio $B_n(x)$ of the torus. Therefore the farest node in a $x$-ball centered at a node $i$ should have a complete rank $NB_n(x)$ for $i$, while being at a distance $x$ from $i$. We deduce that $S_X(x)=S_R(NB_n(x))$. Equation~\eqref{eq:f_sr_k} concludes.
\end{proof}

The value of $B_n(x)$ depends on $n$ and on the norm used. If we conside the maximum norm, then $B_n(x)=\min((2x)^n,1)$. For other norms, the formula may be more complicated because the ball may partially overlap itself in the torus. Note, that if we choose $\mathbb{R}^n$ (with uniform point distribution) instead of the $n$-torus, $B_n(x)$ is just the size of a ball of radius $x$.

\begin{figure}[ht]
	\centering
\includegraphics[width=.5\textwidth]{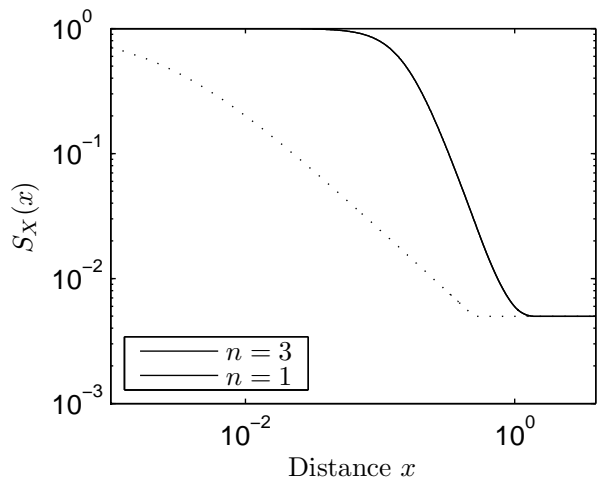}
	\caption{Distance distribution  ($N=2000$,$p=\frac{1}{100}$).}
	\label{fig:geo}
\end{figure}

Figure~\ref{fig:geo} shows $S_X$ for $n=1$ and $n=3$, with the taxicab norm. With this norm, we have $B_1(x)=\min(2x,1)$ and
$$B_3(x)=\left\{ \begin{array}{l}
	\frac{4}{3}x^3\text{ if $0\leq x \leq \frac{1}{2}$,}\\
	\frac{4}{3}x^3-4(x-\frac{1}{2})^3)\text{ if $\frac{1}{2}\leq x \leq 1$,}\\
	1-\frac{4}{3}(\frac{3}{2}-x)^3\text{ if $1\leq x \leq \frac{3}{2}$,}\\
	1\text{ if $x \geq \frac{3}{2}$.}
\end{array} \right.$$
We used $N=2000$ and $p=\frac{1}{100}$, and the fluid limit and empirical distribution of $S_X$ were indistinguishable.

\subsection{Acceptable rank distribution}
\label{sec:relative}

Now we want to investigate the probability that the mate of a node has an acceptable rank $k$. We call $D_r(k)$ this probability. Like for the other distributions, we introduce the CCDF $S_r(k):=1-\sum_{l=1}^{k-1}D_r(k)$.

Following the complete ranking method, we consider the conditions for a node $i$ to be mated with its $k^{th}$ best neighbor $j=r_i^{-1}(k)$:
\begin{itemize}
	\item $i$ must have $k$ neighbors or more,
	\item it must not be mated with someone better than $j$,
	\item $j$ must not be mated with someone better than $i$.
\end{itemize}

With the acceptance ranking, there is intrinsic correlations between these events that complicates things. Despite of that, assuming that these events are independent allows us to give a first, non-accurate, recursive formula:
\begin{equation}
	D_r(k)=S_r(k)\frac{1-I_{1-p}(n-k+1,k)}{k+1}\text{,}
	\label{eq:spetit_r_1}
\end{equation}
where $I_x$ is the regularized incomplete beta function.

\begin{proof} $i$ has $k$ neighbors or more with probability $1-I_{1-p}(n-k+1,k)$. The probability that $i$ is not with better than $j$ is $S_r(k)$. For the reciprocal, we can use $K=\frac{k}{p}$ as a (very rough) approximation of the complete rank; then Equation~\eqref{eq:f_sr_k} gives the probability $\frac{1}{k+1}$. Formula~\eqref{eq:spetit_r_1} follows.
\end{proof}

\begin{figure}[ht]
	\centering
\includegraphics[width=.5\textwidth]{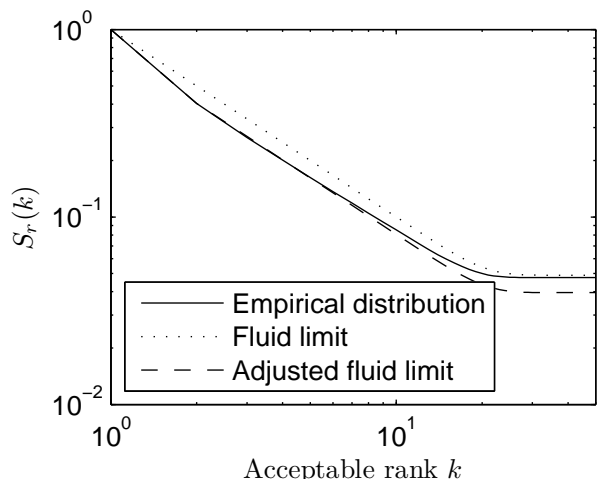}
	\caption{Acceptable ranking CCDF  ($N=2000$,$p=\frac{1}{100}$).}
	\label{fig:accept_ranking}
\end{figure}

The results are shown in Figure~\ref{fig:accept_ranking}. One can observe that Equation~\eqref{eq:spetit_r_1} is not accurate for $D_r(1)$, which provokes a gap between the empirical distribution and the formula.

In an attempt to adjust the formula, we propose a more accurate estimation of $D_r(1)$: under the normalized fluid limit, the scaled rank of the first neighbor $j$ of a given peer $i$ follows the distribution $de^{-d\alpha}$. $j$ and $i$ are mate if $j$ is mated with someone with a scaled rank greater or equal to $\alpha$, which happens with probability $\frac{1}{d\alpha+1}$. Thus we have
\begin{equation}
\begin{array}{r@{ = }l}
D_r(1)& \int_{0}^{\infty}{\frac{d\ed^{-d\alpha}}{d\alpha+1}\dd\alpha}\\
& \int_{1}^{\infty}{\frac{\ed^{-t+1}}{t}\dd t}\\
& \ed E_1(1)\approx 0.596
\end{array}
\end{equation}

\begin{figure}[ht]
	\centering
\includegraphics[width=.5\textwidth]{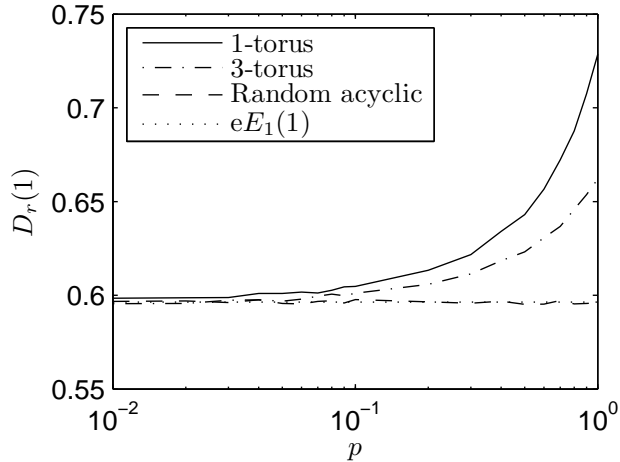}
	\caption{$D_r(1)$ as a function of $p$ ($N=2000$).}
	\label{eq:dr1}
\end{figure}

The accuracy of $D_r(1)=\ed E_1(1)$ ($E_1$ denotes the exponential integral) is verified in Figure~\ref{eq:dr1}. If we use this value for adjusting the fluid limit, we get a better estimation of $S_r$ for small values of $k$ (cf Figure~\ref{fig:accept_ranking}). However, this adjustment introduces a gap for larger values of $k$. In a further version of this paper, we will aim at unifying these two estimates, which will require a better understanding of the correlations that occur when considering the acceptable rank.

\section{\texorpdfstring{$b$}{b}-matching generalization}
\label{sec:bb_mean}

We now extend our results to the case of multiple matchings. For simplicity, we consider here that the quota vector $b$ is a scalar, i.e. that all nodes share the same number of authorized collaborations. For distance and acyclic preferences, we focus on the complete rank, although distance and acceptable ranking could be derived using the same techniques than for $b=1$.

\subsection{Mean Field formulas}

A peer can now have up to $b$ mates. For $1\leq c\leq b$, $D_{c}$ denotes the distribution of the complete ranking of the $c^{th}$ best mate, and $S_{c}$ denotes the corresponding CCDF. Like we did for $b=1$, we can give the conditions for a node $j=R_i^{-1}(K)$ to be the $c^{th}$ mate of a node $i$:
\begin{itemize}
	\item $\{i,j\}$ is an acceptable edge,
	\item the $(c-1)^{th}$ mate of $i$ (if $c>1$) is better than $j$, but the $c^{th}$ (if any) is not,
	\item the $b^{th}$ mate of $j$ (if any) is not better than $i$.
\end{itemize}

By extending Assumption~\ref{ass:meanfield1}, we obtain a generic mean field formula for multiple matchings:
\begin{equation}
	D_{R_i,c}(K)=\left\{
\begin{array}{l}
pS_{R_i,1}(K)S_{R_j,b}(R_j(i)) \text{ if $c=1$, otherwise}\\
p(S_{R_i,c}(K)-S_{R_i,c-1}(K))S_{R_j,b}(R_j(i)) \text{.}
\end{array}
\right.
	\label{eq:generic_meanfield_multiple}
\end{equation}

Like for the simple matching case, this formula can be adapted to specific preferences.

We first consider node-based preferences. $D_c(i,j)$ being the probability that the $c^{th}$ mate of $i$ is $j$, we have the following system, which can be solved by a double iteration on $i$ and $j$ (cf~\cite{DBLP:conf/icdcs/GaiMMR07})
\begin{equation}
	D_c(i,j)=\left\{
\begin{array}{l}
0\text{ if $i=j$,}\\
pS_1(i,j)S_b(j,i)\text{ if $i\neq j$, $c=1$,}\\
p(S_c(i,j)-S_{c-1}(i,j))S_b(j,i)\text{ if $i\neq j$, $c>1$.}
\end{array}
\right.
	\label{eq:nodebase_meanfield_multiple}
\end{equation}

Then, for acyclic and distance-based preferences, we also extend Assumption~\ref{ass:meanfield2} (homogeneity of the distributions and symmetry of the complete ranking). This gives the following system:
\begin{equation}
	D_{R,c}(K)=\left\{\begin{array}{l}
	pS_{R,1}(K)S_{R,b}(K)\text{ if $c=1$,}\\
	p(S_{R,c}(K)-S_{R,c-1}(K))S_{R,b}(K)\text{ if $c>1$.}\\
\end{array} \right.
	\label{eq:drc_meanfield}
\end{equation}

Using $S_{R,c}(1)=1$ and $D_{R_c}(K)=S_{R,c}(K)-S_{R,c}(K+1)$, Equation~\eqref{eq:drc_meanfield} immediately gives an iterative computation of $S_{R,c}$ .

\begin{figure}[ht]
\begin{center}
\subfloat[$b=2$]{ 
\parbox{0.4\textwidth}{\includegraphics[width=.4\textwidth]{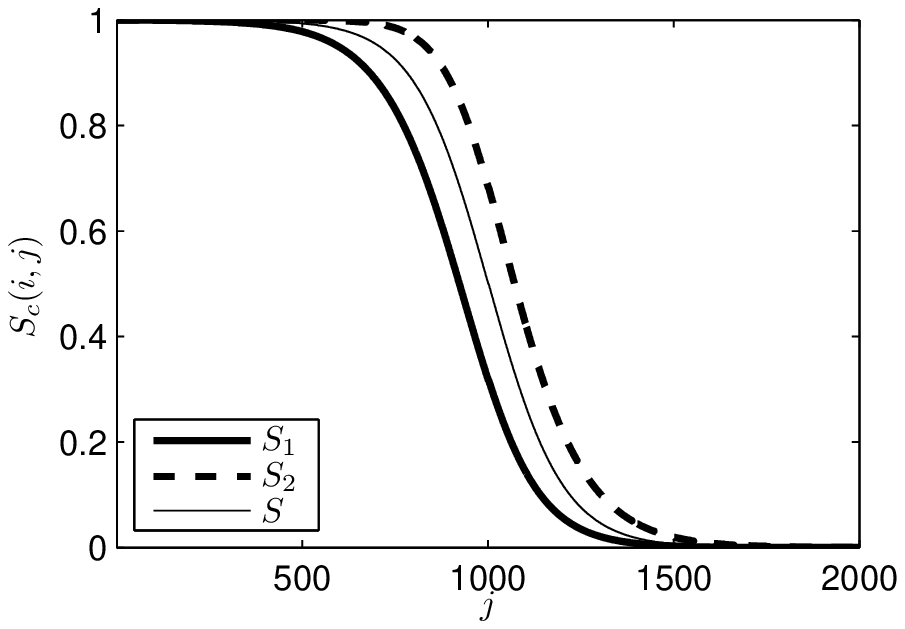}}
\parbox{0.4\textwidth}{\includegraphics[width=.4\textwidth]{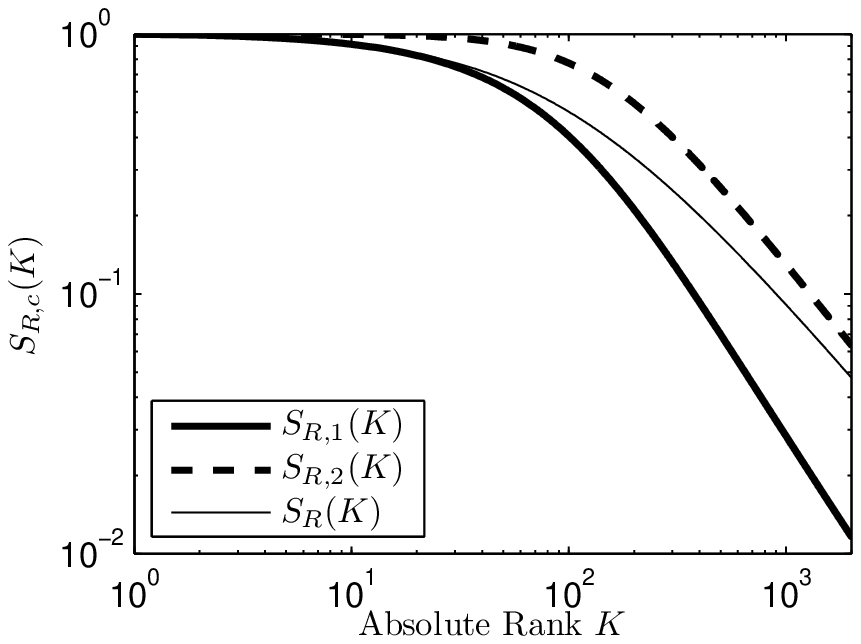}}}\\
\subfloat[$b=3$]{ 
\parbox{0.4\textwidth}{\includegraphics[width=.4\textwidth]{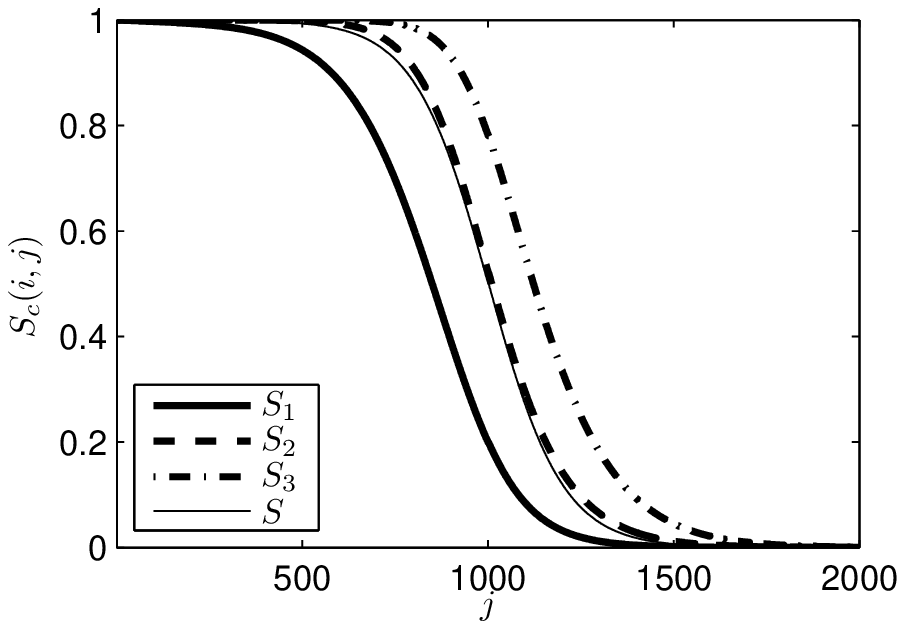}}
\parbox{0.4\textwidth}{\includegraphics[width=.4\textwidth]{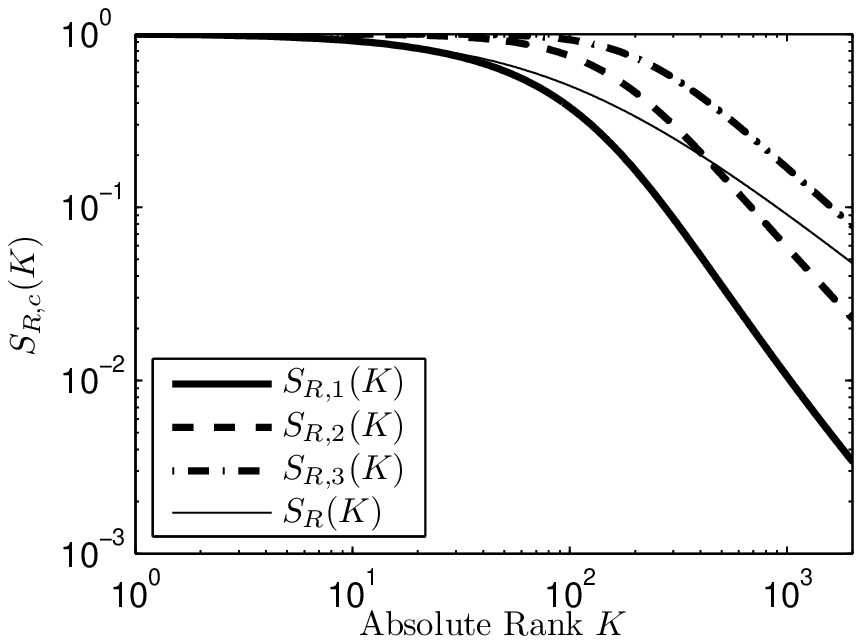}}}\\
\subfloat[$b=4$]{ 
\parbox{0.4\textwidth}{\includegraphics[width=.4\textwidth]{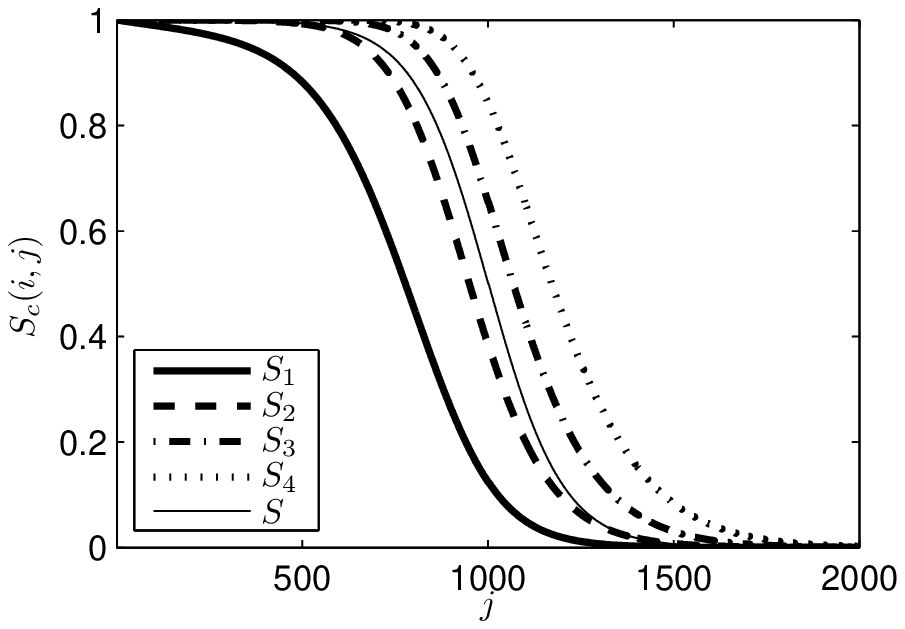}}
\parbox{0.4\textwidth}{\includegraphics[width=.4\textwidth]{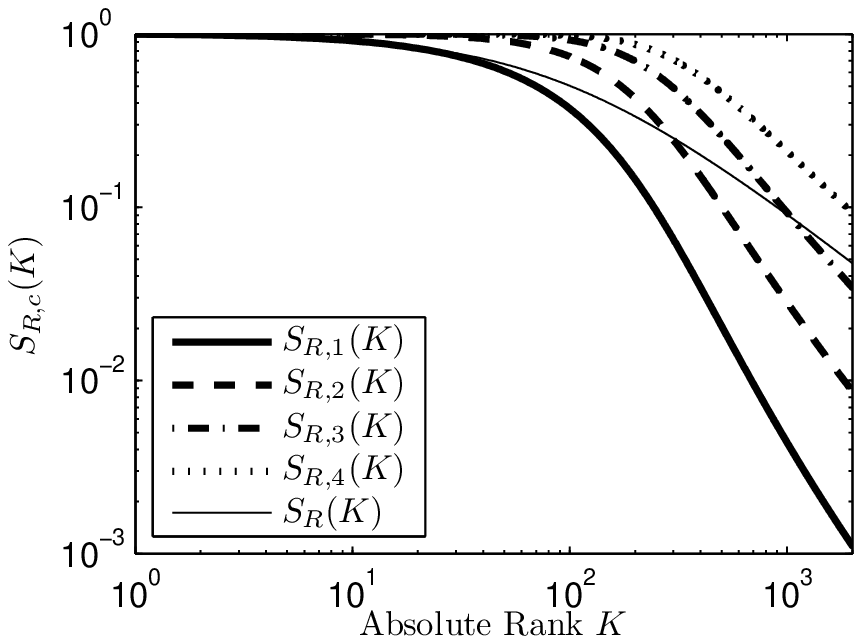}}}\\
	\caption{Complete rank CCDFs for $b>1$. Node-based (resp. acyclic/geometric) distributions are on the left (resp. right) side. $N=2000$, $p=\frac{1}{100}$, and $i=1001$ (for $S_c(i,j)$}
	\label{fig:bvalid}
\end{center}
\end{figure}

Figure~\ref{fig:bvalid} shows $S_c(i,j)$ (node-based) and $S_{R,c}$ (acyclic/geometric) as obtained by \eqref{eq:nodebase_meanfield_multiple} and \eqref{eq:drc_meanfield}. The parameters are $b\in\{2,3,4\}$, $N=2000$, $p=\frac{1}{100}$, and $i=1001$ (for $S_c(i,j)$). We verified for each set of parameters that the curves coincide with the empirical distribution. $S$ and $S_R$ (CCDF for $b=1$) are also plotted for serving as a landmarks. We see that the curves have a behavior that is similar than for the simple matching case: for node-based preferences, it seems that the distribution $D_c(i,.)$ are still exponentially decreasing, even if seems that there is now offsets between the distribution peaks and $i$. For acyclic and geometric preferences, we still observe a kind of power law behavior.

\subsection{Fluid limits}

Fluid limits also exist for $b>1$. We will not present the proofs in this paper, because they are essentially the same that the uniformly Cauchy proofs for the simple matching fluid limits, only more complex to write because of the multiple distribution involved. Therefore we just give the equations verified by the limits.

For node-based preferences, the scaled limit $\mS$ of the CCDF verifies:
\begin{equation}
\partial_y\mS_{c}(\alpha,\beta)=\left\{
\begin{array}{l}
-d\mS_1(\alpha,\beta)\mS_b(\beta,\alpha)\text{ for $c=1$, otherwise}\\
-d\left(\mS_c(\alpha,\beta)-\mS_{c-1}(\alpha,\beta)\right)\mS_b(\beta,\alpha)\text{,}
\end{array}
\right.
\label{eq:bbfluid}	
\end{equation}
with border conditions $\mS_c(\alpha,0)=1$.

Similarly, for acyclic and distance-based preferences, the scaled limit $\mS_R$ of the CCDF verifies
\begin{equation}
\dot{\ms}_{R,c}=\left\{\begin{array}{l}
	-d\ms_{R,1}\ms_{R,b}\text{ if $c=1$,}\\
	-d(\ms_{R,c}-\ms_{R,c-1})\ms_{R,b}\text{ if $c>1$,}\\
\end{array} \right.	
	\label{eq:bfluid}
\end{equation}
\noindent with the boundary condition $\ms_{R,c}(0)=1$.

There is no simple explicit solution for Equations \eqref{eq:bbfluid} and \eqref{eq:bfluid}.
 However,~\eqref{eq:nodebase_meanfield_multiple} and \eqref{eq:drc_meanfield} can still be used as difference equations to approximate a numerical solution. The reason for which we give these limits is that we think that they can give us valuable information about the asymptotical behavior of the distribution (exponentially decreasing or power law), even if this work is still to be done.

\subsection{Discussion}

\subsubsection{Stratification trade-off}

As we have seen, for node-based preferences, the mates of a given peer $i$ have, in average, the same rank than $i$. This is the stratification effect (\cite{DBLP:conf/icdcs/GaiMMR07}), which guarantees a some fairness in the stable configuration: the expected gain of a node tends to be the value offered by this node, measured in term of ranking. However, we also observed that the exponential decreases of the $\mD_c(i,.)$ functions provokes a standard deviation of the same order that $\frac{1}{d}$, where $d$ is the average degree in the acceptance graph. This gives the following stratification trade-off:
\begin{itemize}
	\item if $d$ is too small, the standard deviation is high. In particular, if the mark matrix is non uniformly distributed, there can be a big difference between the expected gain and gift, measured with the marks. This issue has been highlighted in~\cite{DBLP:conf/icdcs/GaiMMR07} for explaining a possible workaround of BitTorrent's Tit-for-Tat policy;
	\item on the other hand, a high $d$ will enforce the fairness. However, the size of the acceptance graph degree has a cost for the nodes (memory usage, overlay management,\ldots). Also, the absence of long-range mates makes the diameter of the stable configuration high, which can be problematic if messages are to be spread using stable edges.
\end{itemize}
Note, that there is a similar trade off for $b$, which is the maximal degree in the stable configuration. This suggests that most node-based preference systems (this includes the systems based on the sharing of an access bandwidth, a storage or CPU capacity, an expected uptime,\ldots) should admit an optimal pair $(d,b)$ with respect to the stable collaborations properties, whose values depend on the weight put on the effects presented above.

\subsubsection{Small-World effect in geometric preferences}
\label{subsec:diaclust}

A small world is a sparse graph with a low average shortest path length (ASPL) and a high clustering coefficient. In details:
\begin{itemize}
	\item \emph{sparse graph} means that the average degree is $O(1)$ or $O(\log n)$,
	\item \emph{low ASPL} means $O(\log(n))$,
	\item \emph{high clustering coefficient} means that two nodes sharing an edge are likely to have a common neighbor. The clustering coefficient is a probability, that must be compared to the clustering coefficient of a random graph with same number of nodes and edges.
\end{itemize}
 In~\cite{kleinberg00small}, Kleinberg proved that a $n$-dimensional grid can be turned into a small world by adding long-range edges that follow a $\Omega(\frac{1}{x^n})$ distribution.

For multiple matchings, the stable configuration in geometric preferences is likely to have a high clustering coefficient, because most of the stable edges link close nodes. Moreover, the power-law rank distribution tells that long-range edges exist. So the stable configuration is likely to be eligible as a small-world.

\begin{table}
\begin{center}
\begin{tabular}{|c|c|c|}
		\hline
	Type of preferences & ASPL & Clustering Coefficient \\
	\hline
	$1$-torus & $7.4$ & $0.055$ \\
	\hline
	$2$-torus & $6.7$ & $0.043$ \\
	 \hline
	 Meridian & $6.1$ & $0.031$ \\
	 \hline
	 $3$-torus & $5.9$ & $0.033$ \\
	\hline
	$4$-torus & $5.1$ & $0.027$ \\
	\hline
	 Random Acyclic & $5$ & $0.0043$\\
	 \hline
\end{tabular}
	\caption{ASPL and clustering ($N=2000$,$p=\frac{1}{10}$,$b=10$).}
	\label{tab:diam_clust}
\end{center}
\end{table}

In Table~\ref{tab:diam_clust}, we give the ASPL and clustering coefficient for some preferences, using the parameters $N=2000$,$p=\frac{1}{10}$,$b=10$. The reference clustering is here $\frac{b}{N-1}\approx 0.005$. We verify that the for the $n$-tori, the stable configurations are small-worlds.
 On the other hand, like previously observed in~\cite{DBLP:conf/europar/GaiLMMRV07}, the stable configurations of random acyclic preferences are not small-worlds, because of their clustering coefficient (they behave like an incomplete $b$-regular graph).

We also calculated the ASPL and clustering obtained by using the Meridian Project's real-world latencies, which are known to produce small-worlds configuration~\cite{DBLP:conf/europar/GaiLMMRV07}. One can observe that the results are very close to the one obtained with the tori. Interestingly, the closest results are those from the $3$-torus, suggesting that somehow, $3$ may be seen as sort of dimension for the latency space. Considering the recent eager for estimating the Internet dimension (see for instance \cite{abrahao08internet}), this unexpected result is appealing: it suggests that the stable configuration, which is only defined by how the nodes rank each other (latencies are used for sorting the nodes, but the actual values are never involved in the construction), could reveal valuable insight about the topology behind a set of distances.

\section{Conclusion}
\label{sec:conclusion}

We gave a statistical description of the stable configurations obtained from node-based preferences, distance-based preferences, and from random acyclic preferences. 
Starting from a generic formula for the rank distribution, we introduced mean-field and fluid limit techniques in order to give explicit formulas. All our results were validated by means of simulations. An interesting consequence of our results is that for distance-based preferences, the stable configurations behave similarly to Kleinberg's grids, and are small-world graphs.

\newpage
\bibliographystyle{abbrv}
\bibliography{RR-6628}

\begin{thebibliography}{10}

\bibitem{DBLP:conf/wine/AbrahamLMO07}
D.~J. Abraham, A.~Levavi, D.~Manlove, and G.~O'Malley.
\newblock The stable roommates problem with globally-ranked pairs.
\newblock In {\em WINE}, volume 4858 of {\em Lecture Notes in Computer
  Science}, pages 431--444. Springer, 2007.

\bibitem{abrahao08internet}
B.~Abrahao and R.~Kleinberg.
\newblock On the internet delay space dimensionality.
\newblock In {\em Proceedings of the 2008 Internet Measurement}, 2008.

\bibitem{gronwall}
R.~Bellman.
\newblock The stability of solutions of linear differential equations.
\newblock {\em Duke Math. J.}, 10:643--647, 1943.

\bibitem{cohen03incentives}
B.~Cohen.
\newblock Incentives build robustness in bittorrent.
\newblock In {\em P2PECON}, 2003.

\bibitem{evans}
L.~C. Evans.
\newblock {\em Partial Differential Equations}.
\newblock American Mathematical Society, 1998.

\bibitem{DBLP:conf/europar/GaiLMMRV07}
A.-T. Gai, D.~Lebedev, F.~Mathieu, F.~de~Montgolfier, J.~Reynier, and
  L.~Viennot.
\newblock Acyclic preference systems in p2p networks.
\newblock In {\em Euro-Par}, volume 4641 of {\em Lecture Notes in Computer
  Science}, pages 825--834. Springer, 2007.

\bibitem{DBLP:conf/icdcs/GaiMMR07}
A.-T. Gai, F.~Mathieu, F.~de~Montgolfier, and J.~Reynier.
\newblock Stratification in p2p networks: Application to bittorrent.
\newblock In {\em ICDCS}, page~30. IEEE Computer Society, 2007.

\bibitem{gale62college}
D.~Gale and L.~Shapley.
\newblock College admissions and the stability of marriage.
\newblock {\em American Mathematical Monthly}, 69:9--15, 1962.

\bibitem{irving00hospitals}
R.~W. Irving, D.~Manlove, and S.~Scott.
\newblock The hospitals/residents problem with ties.
\newblock In {\em SWAT '00: Proceedings of the 7th Scandinavian Workshop on
  Algorithm Theory}, pages 259--271, London, UK, 2000. Springer-Verlag.

\bibitem{irving02stable}
R.~W. Irving and D.~F. Manlove.
\newblock The stable roommates problem with ties.
\newblock {\em J. Algorithms}, 43(1):85--105, 2002.

\bibitem{kleinberg00small}
J.~Kleinberg.
\newblock The small-world phenomenon: an algorithm perspective.
\newblock In {\em STOC '00: Proceedings of the thirty-second annual ACM
  symposium on Theory of computing}, pages 163--170, New York, NY, USA, 2000.
  ACM.

\bibitem{lebedev07using}
D.~Lebedev, F.~Mathieu, L.~Viennot, A.-T. Gai, J.~Reynier, and
  F.~de~Montgolfier.
\newblock On using matching theory to understand {P2P} network design.
\newblock In {\em International Network Optimization Conference (INOC)}, 2007.

\bibitem{flv08}
E.~Lebhar, P.~Fraigniaud, and L.~Viennot.
\newblock The inframetric model for the internet.
\newblock In {\em Proceedings of the 27th IEEE International Conference on
  Computer Communications (INFOCOM)}, pages 1--9, 2008.
\newblock To appear.

\bibitem{marshall93generalizations}
J.~Marshall~Ash, J.~Cohen, C.~Freiling, and D.~Rinne.
\newblock Generalizations of the wave equation.
\newblock {\em Transactions of the American Mathematical Society},
  338(1):57--75, Jul 1993.

\bibitem{DBLP:conf/sss/Mathieu07}
F.~Mathieu.
\newblock Upper bounds for stabilization in acyclic preference-based systems.
\newblock In {\em SSS}, volume 4838 of {\em Lecture Notes in Computer Science},
  pages 372--382. Springer, 2007.

\bibitem{mathieu08self}
F.~Mathieu.
\newblock Self-stabilization in preference-based systems.
\newblock {\em Peer-to-Peer Networking and Applications}, 1(2):104--121, sept
  2008.

\bibitem{mertens05random}
S.~Mertens.
\newblock Random stable matchings.
\newblock {\em J. Stat. Mech.: Theor. Exp.}, page P10008, 2005.

\bibitem{meridianproject}
\protect{Meridian Project}.
\newblock http://www.cs.cornell.edu/People/egs/meridian/.

\bibitem{roth84evolution}
A.~E. Roth.
\newblock The evolution of the labor market for medical interns and residents:
  A case study in game theory.
\newblock {\em Journal of Political Economy}, 92(6):991--1016, 1984.

\bibitem{roth05pairwise}
A.~E. Roth, T.~Sonmez, and M.~Utku~Unver.
\newblock Pairwise kidney exchange.
\newblock {\em Journal of Economic Theory}, 125(2):151--188, December 2005.
\newblock available at
  http://ideas.repec.org/a/eee/jetheo/v125y2005i2p151-188.html.

\bibitem{tan91necessary}
J.~J.~M. Tan.
\newblock A necessary and sufficient condition for the existence of a complete
  stable matching.
\newblock {\em J. Algorithms}, 12(1):154--178, 1991.

\end{thebibliography}

\newpage
\appendix

\section{Proof of Theorem~\ref{th:f_nb_d}}

\label{app:equ_node}

The proof relies on the following steps:
\begin{itemize}
	\item we prove that the $\tilde{\mD_N}$ functions are uniformly Cauchy on $[0,1]^2$;
	\item we use the Cauchy convergence to show that $\mS_N$ and $\tilde{\mD_N}$ have limits $\mS_\infty$ and $\mD_\infty$, and we give a PDE verified by $\mS_\infty$;
	\item we solve the PDE, and use the solution to get $\mD_\infty$.
\end{itemize}

\subsection{Uniform convergence}

Let $N$ be fixed, and $N_1,N_2$ be two integers greater than $N$. The  corresponding Erdös-Rényi probabilities are $p_1=\frac{d}{N_1}$ and $p_2=\frac{d}{N_2}$. We consider the error function defined by 
\begin{equation}
E(\alpha,\beta)=|\mD_{N_1}(\alpha,\beta)-\mD_{N_2}(\alpha,\beta)|\text{.}
	\label{eq:eab}
\end{equation}

For proving that $\tilde{D_N}$ is uniformly Cauchy, we need to find a bound for $E$ that applies for any $(\alpha,\beta)\in[0,1]^2$, and that tends towards $0$ as $N$ goes to infinity.

Let $\alpha_1$, $\beta_1$, $\alpha_2$, $\beta_2$ be respectively $\frac{\lfloor N_1 \alpha \rfloor}{N_1}$, $\frac{\lfloor N_1 \beta \rfloor}{N_1}$, $\frac{\lfloor N_2 \alpha \rfloor}{N_2}$, $\frac{\lfloor N_2 \beta \rfloor}{N_2}$. Using \eqref{eq:dij_sij} and \eqref{eq:sij_sab}, we have, for $k\in\{1,2\}$,

$$
\begin{array}{rl}	\tilde{\mD_{N_k}}(\alpha,\beta)= & d(1-\int_0^{\beta_k}\mD_{N_k}(\alpha_k,x)\ddx)\times\\
&\times(1-\int_0^{\alpha_k}\mD_{N_k}(\beta_k,x)\, \mathrm{d}x)
\end{array} 
$$ 

It would be nice to have $\alpha$ and $\beta$ instead of $\alpha_k$ and $\beta_k$, and $\tilde{\mD}$ instead of $\mD$. In order to do that, we notice the following:
\begin{itemize}
	\item $\mD_{N_k}(\alpha_k,x)=\mD_{N_k}(\alpha,x)$;
	\item same for $\mD_{N_k}(\beta_k,x)$;
	\item $S\leq 1$, so we have $\mD_{N_k}\leq N_kp_k=d$. As $\alpha-\alpha_k<\frac{1}{N_k}\leq\frac{1}{N}$, it follows that 
$\int_{\alpha_k}^{\alpha}\mD_{N_k}(\beta,x)\ddx\leq\frac{d}{N}$;
\item the same with $\alpha$ and $\beta$ switched;
\item $\tilde{\mD}_{N_k}(\alpha,x)$ is bounded by $d$ and only differs from $\mD_{N_k}(\alpha,x)$ for $\lfloor N_k\alpha\rfloor=\lfloor N_kx\rfloor$. It follows that $|\int_0^{\beta}\tilde{\mD}_{N_k}(\alpha,x)-\mD_{N_k}(\alpha,x)|\leq\frac{d}{N}$;
\item the same with $\alpha$ and $\beta$ switched.
\end{itemize}

We deduce that $$\left|\int_0^{\beta_k}\mD_{N_k}(\alpha_k,x)\ddx
- \int_0^{\beta}\tilde{\mD}_{N_k}(\alpha,x)\ddx\right| \leq \frac{2d}{N}\text{,}$$
and the  same with $\alpha$ and $\beta$ switched. Then, if we call
\begin{equation}
\tilde{\mS}_{N_k}(\alpha,\beta)=1-\int_0^{\beta}\tilde{\mD}_{N_k}(\alpha,x)\ddx\text{,}	
	\label{eq:def_stilden}
\end{equation}
we have
\begin{equation}	|\tilde{\mD_{N_k}}(\alpha,\beta)-d\tilde{\mS}_{N_k}(\alpha,\beta)\tilde{\mS}_{N_k}(\beta,\alpha)|\leq \frac{8d^2}{N}\text{.}
	\label{eq:pre_pde}
\end{equation}

This gives us 
\begin{equation}
\begin{array}{rl}
E(\alpha,\beta)\leq & \frac{16d^2}{N}+\\
&d\left|\tilde{\mS}_{N_1}(\alpha,\beta)\tilde{\mS}_{N_1}(\beta,\alpha)-\tilde{\mS}_{N_2}(\alpha,\beta)\tilde{\mS}_{N_2}(\beta,\alpha)\right|
\end{array}	
	\label{eq:cauchy1}
\end{equation}

Using the definition of $\tilde{\mS}_{N_k}$, we see that 

\begin{eqnarray*}
\lefteqn{\left|\tilde{\mS}_{N_1}(\alpha,\beta)\tilde{\mS}_{N_1}(\beta,\alpha)-\tilde{\mS}_{N_2}(\alpha,\beta)\tilde{\mS}_{N_2}(\beta,\alpha)\right|
\leq}\\
& & \int_0^\beta E(\alpha,x)\ddx+\int_0^\alpha E(\beta,x)\ddx+\\
& &\left(\int_0^\beta E(\alpha,x)\ddx\right)\left(\int_0^\alpha E(\beta,x)\ddx\right)	
\end{eqnarray*}

Note, that both $\tilde{\mD}_{N_1}(\alpha,.)$ and $\tilde{\mD}_{N_2}(\alpha,.)$ are probabilities, so we can bound $\int_0^\beta E(\alpha,x)\ddx$ by $2$ in the integral product. Then \eqref{eq:cauchy1} becomes
\begin{equation}
\begin{array}{rl}
E(\alpha,\beta)\leq & \frac{16d^2}{N} +d\left(\int_{0}^{\beta} E(\alpha,x)dx+3\int_{0}^{\alpha} E(\beta,x)dx\right)
\end{array}	
	\label{eq:cauchy3}
\end{equation}

We now want to merge $\alpha$ and $\beta$ into a single variable. Therefore, we define $F(\gamma):=\sup_{
	 \alpha\leq 1,\beta\leq 1,\alpha+\beta\leq \gamma
}E(\alpha,\beta)$. For any $\alpha\leq 1,\beta\leq 1,\gamma\geq\alpha+\beta$, we have
$$
\begin{array}{rl}
	\int_0^{\beta}E(\alpha,x)\ddx & \leq \int_0^{\beta}F(\alpha+x)\ddx \\
	& \leq \int_\alpha^{\alpha+\beta}F(x)\ddx \\
	& \leq \int_0^{\gamma}F(x)\ddx\text{,}
\end{array}
$$
and the same for $\int_0^{\alpha}E(\beta,x)\ddx$. It follows that

\begin{equation}
\begin{array}{rl}
F(\gamma)\leq & \frac{16d^2}{N} +4d\int_{0}^{\gamma} F(x)dx
\end{array}	
	\label{eq:cauchy4}
\end{equation}

It follows that $F(\gamma)\leq \frac{16d^2}{N}e^{4d\gamma}$ by Grönwall's lemma~\cite{gronwall}. As a special case, for all $\alpha,\beta\leq 1$, we have
\begin{equation}
E(\alpha,\beta)\leq F(2)\leq\frac{16d^2}{N}e^{8d}\text{.}	
	\label{eq:bound_for_E}
\end{equation}

This concludes the proof that $\tilde{\mD}_N$ is uniformly Cauchy.

\subsection{PDE}

As  $\tilde{\mD}_N$ is uniformly Cauchy on $[0,1]^2$, it converges towards a function $\mD_\infty$. Using~\eqref{eq:def_stilden}, we deduce that $\tilde{\mS}_N$ converges towards a continuous function $\mS_\infty$, and that $-\mD_\infty$ is the partial derivative of 
$\mS_\infty$ with respect to its second variable.

Then, if we make $N$ go to infinity in \eqref{eq:pre_pde}, we obtain the PDE verified by $\mS_\infty$:
\begin{equation}
\partial_y\mS_\infty(\alpha,\beta)=-d\mS_\infty(\alpha,\beta)\mS_\infty(\beta,\alpha)\text{,}
	\label{eq:PDE_approx}
\end{equation} 
with limit condition $\mS_\infty(\alpha,0)=1$.

Notice that \eqref{eq:PDE_approx} proves that $\mD_\infty$ is continuous.

\subsection{Resolution}

Note, that for $\alpha=0$, \eqref{eq:PDE_approx} immediately gives $\mS_\infty(0,\beta)=e^{-d\beta}$.

To go further, we introduce the auxiliary function $f(\alpha,\beta):=\log(\frac{\mathcal{S_\infty}(\beta,\alpha)}{\mathcal{S}_\infty(\alpha,\beta)})$.

$f$ is skew-symmetric. Its first partial derivative is:
\begin{eqnarray*}
\partial_xf(\alpha,\beta)& = & \frac{\partial_x\mS_\infty(\alpha,\beta)}{\mS(\alpha,\beta)}-\frac{\partial_y\mS_\infty(\beta,\alpha)}{\mS(\beta,\alpha)}\\
& = & \frac{\partial_x\mS_\infty(\alpha,\beta)}{\mS(\alpha,\beta)} + d\mS(\alpha,\beta)
\end{eqnarray*} 

By differentiating again, we get the mixed derivative
\begin{eqnarray*}
	\partial_{xy}f(\alpha,\beta)& = & \frac{\partial_{xy}\mS_{\infty}(\alpha,\beta)}{\mS_\infty(\alpha,\beta)}-\frac{\partial_{x}\mS_\infty(\alpha,\beta)\partial_{y}\mS_\infty(\alpha,\beta)}{\left(\mS_\infty(\alpha,\beta)\right)^2} \\ & & +d\partial_{y}\mS_\infty(\alpha,\beta)\\
	& = & \frac{\partial_{yx}\mS_{\infty}(\alpha,\beta)}{\mS_\infty(\alpha,\beta)}+\frac{\partial_{x}\mS_\infty(\alpha,\beta)d\mS_\infty(\beta,\alpha)}{\mS_\infty(\alpha,\beta)}\\ & &+d\partial_{y}\mS_\infty(\alpha,\beta)\\
	& = & -\frac{d\partial_{x}\mS_\infty(\alpha,\beta)\mS_\infty(\beta,\alpha)}{\mS_\infty(\alpha,\beta)}\\
	&  & -\frac{d\mS_\infty(\alpha,\beta)\partial_{y}\mS_\infty(\beta,\alpha)}{\mS_\infty(\alpha,\beta)}\\
 & & +d\frac{\partial_{x}\mS_\infty(\alpha,\beta)\mS_\infty(\beta,\alpha)}{\mS_\infty(\alpha,\beta)}+d\partial_{y}\mS_\infty(\alpha,\beta)\\
	 & = & 0 \text{ ($=\partial_{yx}f(\alpha,\beta)$)}
\end{eqnarray*}

The only global solutions to the wave equation $f_{xy}=0$ are those of the form $f(\alpha,\beta)=a(\alpha)+b(\beta)$ (see~\cite{marshall93generalizations}, for instance). Given that $f$ is skew-symmetric, the solution is indeed of the form $f(\alpha,\beta)=a(\alpha)-a(\beta)$. The border conditions immediately give $f(\alpha,\beta)=d(\beta-\alpha)$.

We deduce $\mS(\beta,\alpha)$: $\mS(\beta,\alpha)=\mS(\alpha,\beta)e^{d(\beta-\alpha)}$.

If we treat $\mS_\infty$ as a function of $\beta$ with $\alpha$ as parameter, equation~\eqref{eq:PDE_approx} becomes

\begin{equation}
	\dot{\mS_\alpha}(\beta)=-d\mS_\alpha^2(\beta)e^{d(\beta-\alpha)}
	\label{eq:DE_approx}
\end{equation}

From there, one get $\mS_\infty(\alpha,\beta)=\frac{1}{K+e^{d(\beta-\alpha)}}$. Given that $\mS_\infty(\alpha,0)=1$, the solution is:

$$\mS_\infty(\alpha,\beta)=\frac{1}{1-e^{-d\alpha}+e^{d(\beta-\alpha)}}\text{.}$$
Using $\mD_\infty=-\partial_y\mS_\infty$, one get~\eqref{eq:dij_explicit}. This concludes the proof of Theorem \ref{th:f_nb_d}.
\newpage

\section{Exact resolution of the node-based stable configuration}
\label{sec:b1_exact}

\subsection{Recursive formula}

For $b=1$, we can give an explicit recursive formula for $D(i,j)$. The first step is to compute $D(1,k)$, for $2\leq k\leq n$. As $1$ is the best node, it can choose the best of its neighbors, so $D(1,k)$ is the probability that $k$ is the best of $1$'s neighbors. In other words, this is the probability that $k$ is acceptable for $1$, while all nodes $l$ with $1<l<k$ are not. This gives us

\begin{equation}
	D(1,k)=p(1-p)^{k-2}\text{.}
	\label{eq:d1k}
\end{equation}

\begin{figure}[ht]
\begin{center}
\subfloat[$1<k<i$]{\includegraphics[angle=-90,width=.44\textwidth]{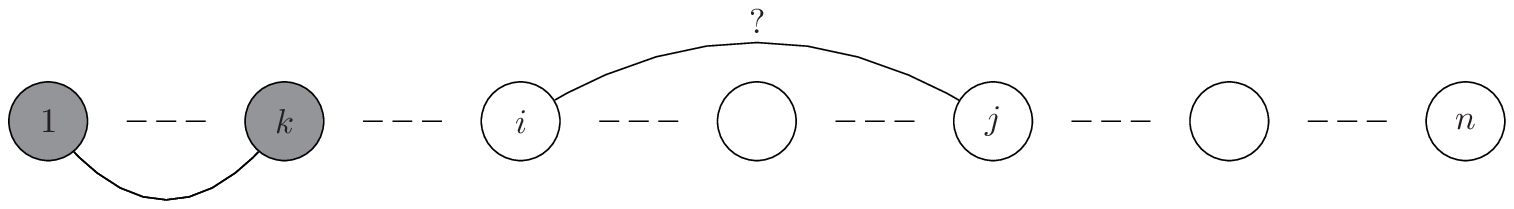}\label{fig:exact1}}\\
\subfloat[$i<k<j$]{\includegraphics[angle=-90,width=.44\textwidth]{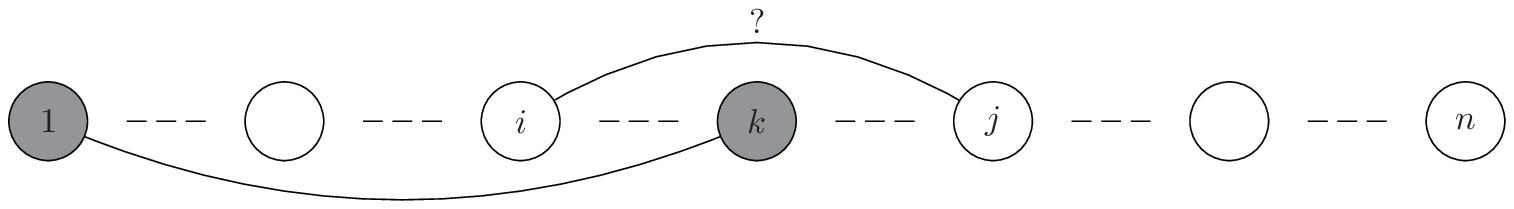}\label{fig:exact2}}\\
\subfloat[$k>j$]{\includegraphics[angle=-90,width=.44\textwidth]{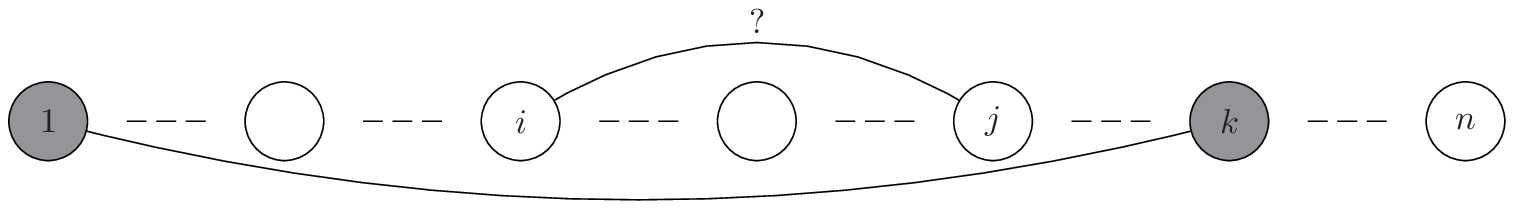}\label{fig:exact3}}
\caption{Using the mate of $1$ to deduce $D(i,j)$.} \label{fig:exact}
\end{center}
\end{figure}

Now, we consider two nodes $i$ and $j$ such that $1<i<j\leq n$. $D(i,j)=P(i\leftrightarrow j)$ can be calculated with a proper conditionning on the mate $k$ (if any) of $1$. The key is to notice that if $1$ is mated with $k$, the both of them can be virtually removed from the graph. The remaining graph is still Erdös-Rényi and the probabilities are the same up to a slight relabeling:
\begin{itemize}
	\item if $k=i$ or $k=j$, then $i$ cannot be mated with $j$;
	\item if $1<k<i$, $i$ and $j$ can be virtually relabeled $i-2$ and $j-2$ (cf Figure~\ref{fig:exact1}), so we have $P(i\leftrightarrow j|1<k<i)=P((i-2)\leftrightarrow (j-2))$;
	\item if $i<k<j$, $i$ and $j$ can be virtually relabeled $i-1$ and $j-2$ (cf Figure~\ref{fig:exact2}), so we have $P(i\leftrightarrow j|i<k<j)=P((i-1)\leftrightarrow (j-2))$;
	\item if $1$ is not mated or $k>j$ (notation: $k \nleq j$), $i$ and $j$ can be virtually relabeled $i-1$ and $j-1$ (cf Figure~\ref{fig:exact3}), so we have $P(i\leftrightarrow j|k \nleq j)=P((i-1)\leftrightarrow (j-1))$.
\end{itemize}

Under this conditioning, we get
\begin{equation}
\begin{array}{rl}
D(i,j)= & P(i\leftrightarrow j|1<k<i)P(1<k<i)\\
& +P(i\leftrightarrow j|i<k<j)P(i<k<j)\\
& +P(i\leftrightarrow j|k\nleq j)P(k\nleq j)\text{.}
\end{array}
\label{eq:dij_condi}
\end{equation}
 
This leads to the following formula for $D$:
\begin{equation}
\begin{array}{rl}
D(i,j)= & \left(1-S(1,i)\right)D(i-2,j-2)\\
&+\left(S(1,i+1)-S(1,j)\right)D(i-1,j-2)\\
&+S(1,j+1)D(i-1,j-1)
\text{.}
\end{array}	
	\label{eq:dij_exact}
\end{equation}

From $\eqref{eq:d1k}$, we have $S(1,k)=(1-p)^{k-2}$. This gives
\begin{equation}
\begin{array}{rl}
D(i,j)= & A(i)D(i-2,j-2)+B(i,j)D(i-1,j-2)+C(j)D(i-1,j-1)
\text{, with}\\
A(i)= & 1-(1-p)^{i-2}\\
B(i,j) = & (1-p)^{i-1}-(1-p)^{j-2}\\
C(j) = & (1-p)^{j-1}
\end{array}	
	\label{eq:dij_exact2}
\end{equation}

Now, in order to give a fluid limit, it can be convenient to reduce \ref{eq:dij_exact} to an expression of the complementary cumulative distribution $S$. Using the definition $S(i,j)=\sum_{l\nless j}D(i,l)$, Equation \ref{eq:dij_exact2} becomes, after simplification,

\begin{equation}
\begin{array}{rl}
S(i,j)= & A(i)S(i-2,j-2)+B(i,j)S(i-1,j-2)+C'(j)S(i-1,j-1)\text{, with}\\
C_2(j) = & (1-p)^{j-2}\text{.}	
\end{array}	
	\label{eq:sij_exact}
\end{equation}

\subsection{Uniform convergence}

Like for the mean formula, we can prove that the scaling $\mD_N(\alpha,\beta)=ND(\atoi{N\alpha}+1,\atoi{N\beta}+1)$ is uniformly Cauchy . The sketch of proof is the same: clean the boundary of the integrals and the other $O(\frac{1}{N})$ offsets, then use an auxiliary error variable $\gamma$ and use Grönwall's lemma to conclude. This guarantees the convergence of $\mD_N$ and $\mS_N$.

\subsection{PDE}

We will use the fact that if we use the scaling $i\rightarrow \left\lfloor N\alpha \right\rfloor+1$, $j\rightarrow \left\lfloor N\beta \right\rfloor+1$, then
\begin{itemize}
	\item $A(i)$ converges towards $1-e^{-d\alpha}$,
	\item $B(i,j)$  converges towards $e^{-d\alpha}-e^{-d\beta}$,
	\item $C(j)$ and $C_2(j)$ both converge towards $e^{-d\beta}$.
\end{itemize}

The first step is to \emph{translate}~\eqref{eq:sij_exact} into an expression of $\mS_N$: with $\alpha=\frac{i-1}{N}$ and $\beta=\frac{j-1}{N}$, we obtain

\begin{equation}
\begin{array}{rl}
\mS_N(\alpha,\beta)= & A(i)\mS_N(\alpha-\frac{2}{N},\beta-\frac{2}{N})+B(i,j)\mS_N(\alpha-\frac{1}{N},\beta-\frac{2}{N})+C_2(j)\mS_N(\alpha-\frac{1}{N},\beta-\frac{1}{N})\text{.}	
\end{array}	
	\label{eq:srondeij_exact}
\end{equation}

We notice that $(A+B+C)(i,j)=1-p(1-p)^{i-2}$. 
If we remove $(A+B+C)(i,j)\mS_N(\alpha,\beta-\frac{2}{N})$ from each side of \eqref{eq:sij_exact}, and multiply the result by $N$:
\begin{itemize}
	\item the left part becomes
$$N(\mS_N(\alpha,\beta)-\mS_N(\alpha,\beta-\frac{2}{N}))+d(1-p)^{i-2}\mS_N(\alpha,\beta-\frac{2}{N})\text{,}$$
which converges as $N\rightarrow \infty$ towards

$$2\frac{\partial \mathcal{S}_\infty}{\partial \beta}+de^{-d\alpha}\mathcal{S}_\infty\text{,}$$

\item the right part becomes

$$
\begin{array}{rl}
A(i)N(\mS_N(\alpha-\frac{2}{N},\beta-\frac{2}{N})-\mS_N(\alpha,\beta-\frac{2}{N}))&+(B(i,j)+C(j))N(\mS_N(\alpha-\frac{1}{N},\beta-\frac{2}{N})-\mS_N(\alpha,\beta-\frac{2}{N}))\\
&+C(j)N(\mS_N(\alpha-\frac{1}{N},\beta-\frac{1}{N})-\mS_N(\alpha-\frac{1}{N},\beta-\frac{2}{N}))\\
\end{array}\text{,}$$
which converges as $N\rightarrow \infty$ towards

$$-2(1-e^{-d\alpha})\frac{\partial \mathcal{S}_\infty}{\partial \alpha}-e^{-d\alpha}\frac{\partial \mathcal{S}_\infty}{\partial \alpha}+e^{-d\beta}\frac{\partial \mathcal{S}_\infty}{\partial \beta}\text{.}$$

\end{itemize}

So after scaling, the recursive equation is now:

$$2\frac{\partial \mathcal{S}_\infty}{\partial \beta}+de^{-d\alpha}\mathcal{S}_\infty=-2(1-e^{-d\alpha})\frac{\partial \mathcal{S}_\infty}{\partial \alpha}-e^{-d\alpha}\frac{\partial \mathcal{S}_\infty}{\partial \alpha}+e^{-d\beta}\frac{\partial \mathcal{S}_\infty}{\partial \beta}\text{.}$$

In other words, $\mathcal{S}$ verifies the PDE:

$$(2-e^{-d\alpha})\frac{\partial \mathcal{S}_\infty}{\partial \alpha}+(2-e^{-d\beta})\frac{\partial \mathcal{S}_\infty}{\partial \beta}+de^{-d\alpha}\mathcal{S}_\infty=0\text{.}$$

\begin{theorem}
With the border condition $\mathcal{S}_\infty(0,\beta)=e^{-d\beta}$, the unique solution of this PDE is $\mathcal{S}_\infty=\frac{1}{1-e^{-d\alpha}+e^{d(\beta-\alpha)}}$.

The scaled version of $D$, denoted $\mathcal{D}_\infty$, thus verifies:

$$\mathcal{D}_\infty(\alpha,\beta)=-\frac{\partial \mathcal{S}_\infty}{\partial \beta}(\alpha,\beta)=\frac{de^{d(\beta-\alpha)}}{(1-e^{-d\alpha}+e^{d(\beta-\alpha)})^2}\text{.}$$
\end{theorem}

\begin{proof}
Let us change the variables: put $x=e^{d \alpha}$ and $y=e^{d \beta}$. We also make the PDE more symmetric by multiplying by $x$. Define $u$ by putting $\mathcal{S}_\infty(\alpha,\beta)= x u(x,y)$.
The PDE then becomes: 

$$(2-\frac{1}{x})(d x^2 \frac{\partial u}{\partial x}+dx u)+(2-\frac{1}{y}) x d y \frac{\partial u}{\partial y}+d\frac{1}{x} x u=0\text{.}$$

ie:

$$\left\{
\begin{matrix}
(2x-1)\frac{\partial u}{\partial x}+(2y-1) \frac{\partial u}{\partial y}+2 u=0\\
u(1,y)=\frac{1}{y}
\end{matrix}
\right.
$$

This equation is a non-linear first order PDE: $F(Du,u,x)=0$, where $F$ is linear. To solve this PDE, we use the classical method of \emph{characteristics} described in \cite{evans}, chapter 3. Let $X(s)=(x(x),y(s))$ ($s$ in an interval of $\mathbb{R}$), be a trajectory in the base space; define $p(s)=Du(X(s))$ and $z(s)=u(X(s))$. Then, solving the equation $F(p(s),z(s),S(s))=0$ leads to the equivalent system of ODE (we forget about $p(s)$, which is not required to solve the PDE with boundary condition, see \cite{evans} p 100 for further precisions):

$$\left\{
\begin{matrix}
\dot{x}(s)=2x(s)-1\\
\dot{y}(s)=2y(s)-1\\
\dot{z}(s)=-2z(s)\\
z_0=z(x_0:=1,y_0)=\frac{1}{y_0}
\end{matrix}
\right.
$$

where $\dot{\ }$ stands for $\frac{d}{ds}$.

These 3 ODEs are with separable variables (Cauchy-Lipschitz theorem applies for existence and unicity). The solution with the boundary condition at $s=0$, $x_0=1$, $y_0\in\mathbb{R}$, $z_0=\frac{1}{y_0}$ is:

$$\left\{
\begin{matrix}
2x(s)-1=e^{2s}\\
2y(s)-1=(2y_0-1) e^{2s}\\
z(s)=\frac{1}{y_0}e^{-2s}\\
\end{matrix}
\right.$$

Now given $(x,y)$, we deduce $s$ such that $x(s)=x$ and $y(s)=y$ then $y_0$ and $z(s)=u(x,y)$:
$2y_0-1=\frac{2y-1}{2x-1}$ then

$$u(x,y)=\frac{1}{y_0}{\frac{1}{2x-1}}=\frac{1}{x+y-1}\text{.}$$

Replacing $x$ and $y$ by $e^{d\alpha}$ and $e^{d\beta}$ concludes the proof.
\end{proof}

\end{document}